\documentclass[%
 reprint,
bibnotes,
 amsmath,amssymb,
 aps,
]{revtex4-1}

\usepackage[pdftex]{graphicx}
\usepackage{dcolumn}
\usepackage{bm}
\usepackage{theorem}
\theorembodyfont{\normalfont}

\usepackage{color}
\usepackage{comment}
\usepackage{txfonts}

\newcommand{\argmax}{\mathop{\rm arg~max}\limits}

\newif\iffigure
\figuretrue

\begin{document}

\preprint{APS/123-QED}

\title{Bayesian Hamiltonian Selection in X-ray Photoelectron Spectroscopy}

\author{Yoh-ichi Mototake$^1$}
\author{Masaichiro Mizumaki$^2$}
\author{Ichiro Akai$^{3,4}$}
\author{Masato Okada$^{1,5}$}
\affiliation{%
Graduate School of Frontier Sciences, The University of Tokyo,  Kashiwa, Chiba 277-8561, Japan$^1$\\
Japan Synchrotron Radiation Research Institute (JASRI), 1-1-1, Kouto, Sayo-cho, Sayo-gun, Hyogo 679-5198, Japan$^2$\\
Institute of Pulsed Power Science, Kumamoto University, Kumamoto 860-8555, Japan$^3$\\
Kyushu Synchrotron Light Research Center, Tosu, Saga 841-0005, Japan$^4$\\
Research and Services Division of Materials Data and Integrated Systems, National Institute for Materials Science, Sengen, Tsukuba, Ibaraki 305-0047, Japan$^5$
}%




\date{\today}

\begin{abstract}
Core-level X-ray photoelectron spectroscopy (XPS) is a useful measurement technique for investigating the electronic states of a strongly correlated electron system. 
Usually, to extract physical information of a target object from a core-level XPS spectrum, 
we need to set an effective Hamiltonian by physical consideration so as to express complicated electron-to-electron interactions in the transition of core-level XPS, and manually tune the physical parameters of the effective Hamiltonian so as to represent the XPS spectrum. Then, we can extract physical information from the tuned parameters. 
In this paper, we propose an automated method for analyzing core-level XPS spectra based on the Bayesian model selection framework, which selects the effective Hamiltonian and estimates its parameters automatically. 
The Bayesian model selection, which often has a large computational cost, was carried out by the exchange Monte Carlo sampling method. By applying our proposed method to the 3$d$ core-level XPS spectra of Ce and La compounds, we confirmed that our proposed method selected an effective Hamiltonian and estimated its parameters appropriately; these results were consistent with conventional knowledge obtained from physical studies. Moreover, using our proposed method, we can also evaluate the uncertainty of its estimation values and clarify why the effective Hamiltonian was selected. Such information is difficult to obtain by the conventional analysis method.
\end{abstract}

\pacs{Valid PACS appear here}
\maketitle



\section{Introduction}
Core-level X-ray photoelectron spectroscopy (XPS) is a useful means of investigating the electronic state of a strongly correlated electron system\cite{kanamori87}. 
The XPS spectral distribution is reproduced from its state transition probability. The state transition probability is calculated on the basis of an effective Hamiltonian, which explains complex electron-to-electron interactions in the transition of XPS. 
The parameters of the effective Hamiltonian are the physical parameters of the measurement target. 
Therefore, by adjusting the parameters of the effective Hamiltonian so as to reproduce the measured XPS spectrum, the physical parameters of the object material are estimated as the tuned parameters\cite{Kotani74,Kotani85,Kotani87,Groot08}. 
Normally, this parameter adjustment is performed manually, and the effective Hamiltonian, which is the premise of the analysis, is set on the basis of the physical consideration of the XPS transition process by researchers\cite{Kotani74,Kotani85,Kotani87,Groot08}.
\par
There is an information science methodology called spectral deconvolution that regresses a spectrum as a linear sum of unimodal basis functions, such as the Gaussian function. 
Nagata et al.\cite{Nagata12} introduced Bayesian inference to spectral deconvolution by the exchange Monte Carlo method,\cite{Hukushima96} which is an efficient sampling method. 
By Bayesian spectral deconvolution, the quantitative selection of the spectral model, such as the selection of the number of peaks, has been realized. 
Bayesian spectral deconvolution has been further developed, and the efficient estimation of noise intensity\cite{tokuda16},  the analysis of time series spectra\cite{murata16}, and a fast Bayesian spectral deconvolution algorithm\cite{mototake18} applicable to a high-dimensional spectrum have been realized. 
Many studies have used Bayesian spectral deconvolution to analyze spectra measured in various scientific fields\cite{kasai16,iwamitsu16,hong16}. 
Normally, in spectral deconvolution, the parameters of the unimodal basis, such as the position, intensity, and variance of basis functions, 
are estimated. 
Therefore, if the parameters of the physical model do not directly correspond to the parameters of the unimodal basis function, we need an indirect estimation of the parameters. 
In the analysis of the core-level XPS spectrum, the parameters of the effective Hamiltonian also do not directly correspond to the peak positions or peak intensities. 
To realize the direct estimation of 
physical parameters, Bayesian spectral deconvolution methods that build an internal model on the spectrum model have been developed\cite{kasai16,murata16}. 
Here, the internal model is defined as a model whose parameters are physical parameters themselves, unlike the parameters of the unimodal basis function such as the mean, intensity, and variance. 
For instance, Kasai et al. applied the relationship between the protein species and the peak intensity ratio as an internal model, 
and estimated the protein species directly\cite{kasai16}.  
Also, Murata et al. applied a probability differential equation representing the time evolution of peak positions of the spectrum, 
and estimated the parameters of latent dynamics directly\cite{murata16}. 
\par
In this study, we aim to achieve the automatic selection of an effective Hamiltonian and the estimation of its parameters based on the support of statistical science. 
By incorporationg the effective Hamiltonian of core-level XPS into the Bayesian spectral deconvolution method as an internal model, we propose a Bayesian spectral deconvolution method that enables the automatic analysis of core-level XPS spectra. 
Our proposed method is not merely an automated spectral analysis method for estimating the parameters and selecting a model. 
In Bayesian inference, a physical parameter is treated as a statistical variable. 
Therefore, the physical parameter is estimated as the probability distribution of 
the values that the parameter can possibly take, which is called posterior distribution. 
As a result, we can estimate the parameter and, at the same time, determine its estimation accuracy. 
This makes it possible to obtain the information necessary for developing measurement plans, such as the number of measurements and the measurement method, to satisfy the required estimation accuracy. 
In addition, by analyzing the shape of the distribution, we can also discuss the nature of the effective model. 
For example, we can discuss the uncertainty of the model parameters when fitting the model to the data. 
This provides information such as the relationship between the expressive ability of the model and the data complexity. 
It is difficult to obtain this information from simple parameter fitting. 
Thus, our proposed method can extract useful information for physical discussion from XPS spectra. 
\par
In the current methods of analyzing core-level XPS spectra, researchers use effective Hamiltonians such as the cluster model\cite{groot90} and the impurity Anderson model\cite{gunnarsson83, jo88}. 
The cluster model contains the whole process of core-level XPS spectrum generation, such as the spin-orbit interaction, multiplet effect, and crystal field effect. 
In the impurity Anderson model, the effect of the conduction band structure is considered on the hybridization interaction between 4$f$ electrons and conduction electrons. 
In contrast, as the internal model, our proposed method adopts two of the simplest effective Hamiltonians that can represent the XPS spectra of La$_2$O$_3$ and CeO$_2$. The analysis of the core-level XPS spectra started from these effective Hamiltonians\cite{Kotani74, Kotani85}. 
Since these effective Hamiltonians do not contain many factors, such as the effects of the spin-orbit interaction, multiplet, and crystal field, 
they cannot be applied to the general core-level XPS spectra of a rare-earth insulator compound. 
However, the framework of the proposed method does not limit the inner model to these effective Hamiltonians. 
Even if the effective Hamiltonian is replaced with an arbitrary one in the framework of the proposed method, this simply increases the number of parameters to be estimated and does not change the principle of the framework. 
This suggests that our proposed method is applicable to general core-level XPS spectra. 
\par
Finally, we applied the proposed method to emulated measurement data of La$_2$O$_3$ and CeO$_2$ and their intermediate electron states. 
As a result, it was confirmed that the two effective Hamiltonians, which are also regarded as effective Hamiltonians in conventional physical studies\cite{Kotani74, Kotani85}, were selected for the spectrum data of La$_2$O$_3$ and CeO$_2$. 
Furthermore, from the analysis of the posterior distribution, it was shown that the effective Hamiltonian of CeO$_2$ has a too large degree of freedom to express the spectrum data of La$_2$O$_3$. 
In addition, we also applied our proposed method to the spectra, which are the intermediate states of a seamless transition from the CeO$_2$ spectrum (three peaks) to the La$_2$O$_3$ spectrum (two peaks). 
As a result, it was confirmed that the proposed method selects an effective Hamiltonian on the basis of not only the information about the peak number but also other information contained in the effective Hamiltonian. 
This cannot be realized by existing spectral deconvolution methods applicable to XPS spectra, which select the model on the basis of the peak number\cite{Nagata12,tokuda16}. 
%
%

In this way, it was shown that the proposed method can be used to select an effective Hamiltonian that is consistent with physical knowledge. 
It was also shown that, unlike the simple parameter fitting, the proposed method gives us the knowledge necessary for the discussion of physics. 

\section{Generative Model of Spectrum}
\label{gen_model}
In this study, we analyzed spectra that were simulated on the basis of an effective model that is applicable to emulating all types of simplified 4$f$ electron derived 3$d$ core-level XPS spectra of rare-earth insulating compounds. 
We refer to this effective model as a generative model. 
The generative model was proposed by Kotani et al.\cite{Kotani85} 
It employs one of the simplest cluster models 
as the effective Hamiltonian. 
In the effective Hamiltonian, the effect of the spin-orbit interaction, the multiplet effect, and the crystal field effect are ignored. 
Moreover, the effect of hybridization between 4$f$ electrons and conduction electrons is simplified because it assumes that there is only one ligand state. 
Note here that the generative model contains two effective models, which we applied later as the inner model of the spectral deconvolution method. 
We refer to these effective models as recognition models,  
which are explained later. 
\par
There are three eigenstates of $4f$ electrons: 
$|f^0\rangle{}, |f^1\rangle{}$, and $|f^2\rangle{}$. 
In the $|f^0\rangle{}$ state, there is no electron in the $4f$ orbital, whereas there is one electron in the $|f^1\rangle{}$ state and two in the $|f^2\rangle{}$ state. 
In such a state space of 4$f$ electrons, the effective Hamiltonian of the generative model is given as 
\begin{equation}
\begin{split}
 \mathcal{H} &= \epsilon_L \sum_{\nu}  a^{\dagger}_{L\nu}a_{L\nu} + \epsilon^0_f \sum_{\nu} a^{\dagger}_{f\nu}a_{f\nu} + \epsilon_c a^{\dagger}_c a_c\\
&+ \frac{V}{\sqrt{N_f}} \sum_{\nu}  (a^{\dagger}_{f\nu}a_{f\nu} + a_{f\nu}a^{\dagger}_{f\nu}) + U_{ff} \sum_{\nu\rangle{}\nu'} a^{\dagger}_{f\nu}a_{f\nu}a^{\dagger}_{f\nu'}a_{f\nu'}\\
&-U_{fc} \sum_{\nu} a^{\dagger}_{f\nu}a_{f\nu}(1- a^{\dagger}_c a_c),
\end{split}
\end{equation}
where $\epsilon_L$, $\epsilon^0_f$, and $\epsilon_c$ are the energies of the conducting electron of $4f$ rare-earth metals ($5d$, $6s$ electrons), the $4f$ electron, and the core electron, respectively. 
The index $\nu$ $(\nu=1 \textendash N_f, N_f=14)$ represents the quantum number of the spin and $f$-orbital. 
$V$, $U_{ff}$, and $-U_{fc}$ are the energies of the hybridization interaction between 4$f$ electrons and conduction electrons, the Coulomb interaction between $4f$ electrons, and the core-hole Coulomb potential for $4f$ electrons, respectively. 
In both the initial and final states, we set the energy level of $|f^0\rangle{}$ to 0 as the standard energy level. Then, the number of parameters of the effective Hamiltonian $\mathcal{H}$ is reduced to five: $\Delta (= \epsilon^0_f - \epsilon_L), V ,U_{ff}, U_{fc}$, and $\Gamma $. 
For later use, we define the parameter set of the effective Hamiltonian $\mathcal{H}$ as $\boldsymbol{\vartheta} = \{\Delta,V,U_{ff},U_{fc},\Gamma\}$. 
\par
\begin{figure}[t]
  \begin{center}
    \includegraphics[clip,width=7.0cm]{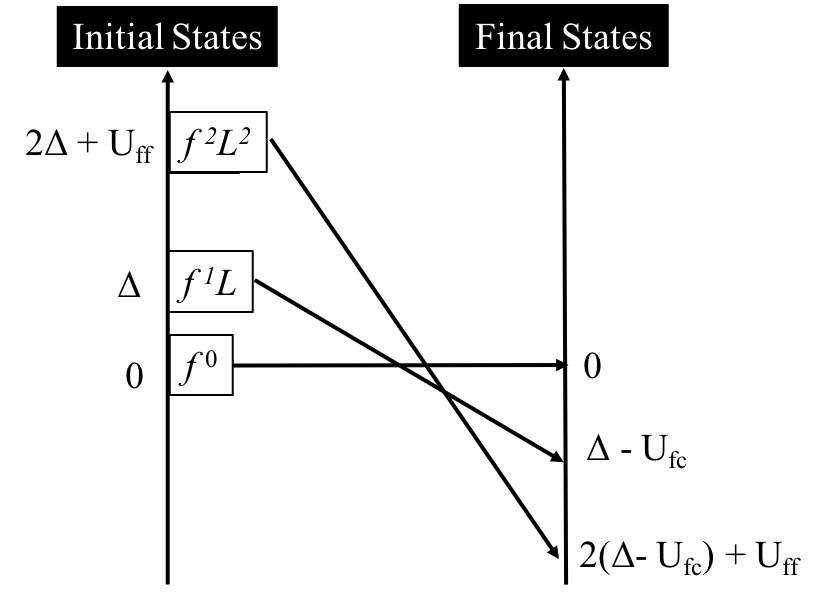}
    \caption{Conceptual figure of the initial and final states of the XPS process. Each state represents the state of the 4$f$ electron $|f^j\rangle{}$$(j=0,1,2)$.}
    \label{fig_model}
  \end{center}
\end{figure}
\begin{figure*}[t]
  \begin{center}
    \includegraphics[clip,width=16.0cm]{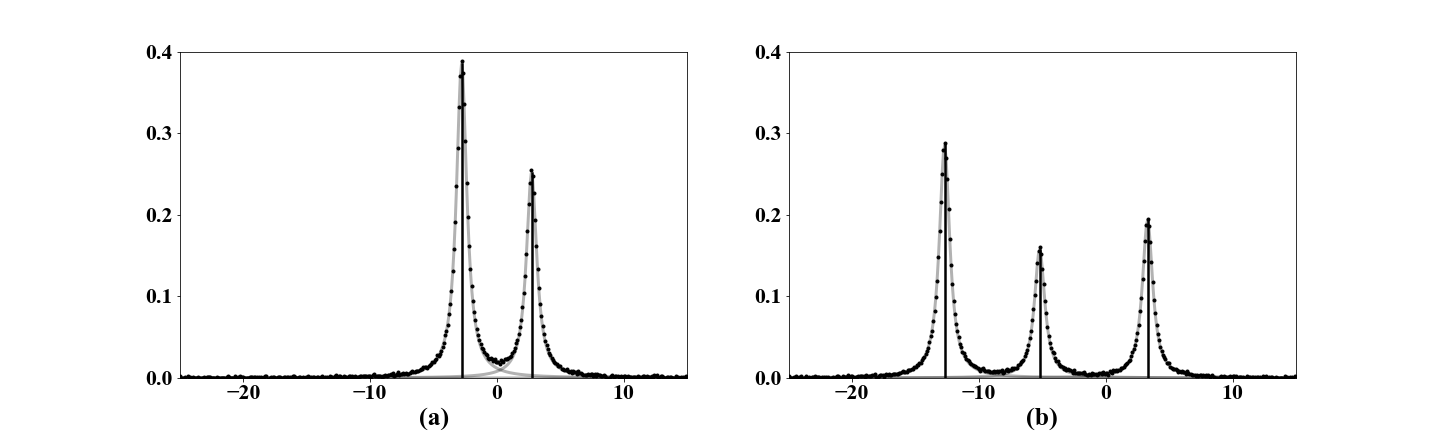}
    \caption{(a) XPS spectrum of La$_2$O$_3$. Gray dots represent the observed data, and the curved gray line represents the true spectral density, the black line represents the peak position and peak intensity. (b) XPS spectrum of CeO$_2$. Both spectra include Gaussian noise with a standard deviation of $0.001$．}
    \label{fig_spectraldata}
  \end{center}
\end{figure*}

The conceptual diagram of the $4f$ electron state from the initial state to the final state is shown in Fig. \ref{fig_model}. 
The initial state is defined as the state before X-rays are irradiated, 
and the final state is defined as the state after X-rays are irradiated and a core hole is generated. 
Furthermore, we define the initial eigenstate of the minimum energy $E_G(\boldsymbol{\vartheta})$ as $|G\rangle{}$ and the final eigenstates of the three energy levels $E_j(\boldsymbol{\vartheta})$ $(j = 0,1,2)$  as $|F_j\rangle{}$ $(j = 0,1,2)$ (Fig. \ref{fig_model}). 
Using Fermi's golden rule, 
we define the transition probability from 
the initial ground state to the final state as above
\begin{equation}
F(\omega;\boldsymbol{\vartheta}) = \sum_{j=0}^2 |\langle{}F_j|a_c|G\rangle{}|^2\delta (\omega - E_j(\boldsymbol{\vartheta}) + E_G(\boldsymbol{\vartheta})).
\end{equation}
By convoluting $F(\omega;\boldsymbol{\vartheta})$ with the Lorentz function 
and by adding the Gaussian noise $\epsilon$, 
we can obtain the intensity distribution of the spectrum as 
\begin{equation}
\begin{split}
\mathcal{I}(\omega;\boldsymbol{\vartheta}) 
=\sum_{j=0}^2 |\langle{}F_j|a_c|G\rangle{}|^2 \frac{\Gamma/\pi}{(\omega-(E_j(\boldsymbol{\vartheta})-E_G(\boldsymbol{\vartheta})))^2+\Gamma^2} + \epsilon,
\end{split}
\end{equation}
where $\Gamma$ is one-half the width of the Lorentz function. 
We define the data generated following $\mathcal{I}(\omega;\boldsymbol{\vartheta})$ as ``emulated measurement data". 
Because the generative model is uniquely defined by the effective Hamiltonian $\mathcal{H}$, we label this model as $\mathcal{H}$. 
\begin{table}[b]
  \caption{Reproduction parameters of XPS spectra of La$_2$O$_3$ and CeO$_2$.}
\begin{center}
  \begin{tabular}{c||c|c|c|c|c} \hline
    Parameter & $\Delta$ &$V$ & $U_{ff}$ & $U_{fc}$ & $\Gamma$ \\ \hline \hline
    La$_2$O$_3$ & {\bf 12.5}  & 0.57 & 10.5 & 12.7 & 0.5    \\ \hline 
    CeO$_2$ & {\bf 1.6} & 0.76 & 10.5 & 12.5 & 0.7  \\ \hline  
  \end{tabular}
  \label{table1}
\end{center}
\end{table}
Following Groot and Kotani\cite{Groot08}, the emulated parameters of the XPS spectrum of La$_2$O$_3$ and CeO$_2$ are set as illustrated in Table \ref{table1}. 
The generated data based on Table \ref{table1} are shown in 
Fig. \ref{fig_spectraldata}. 
The two peaks in Fig. \ref{fig_spectraldata}(a) correspond to La$_2$O$_3$ 
and the three peaks in Fig. \ref{fig_spectraldata}(b) correspond to CeO$_2$. 
\par
In the parameters of the two-peak spectrum, 
the peak intensity of the highest energy $E_2$ 
becomes almost zero. 
This is caused by the decreased interaction between the $|f^2\rangle{}$ state 
and the $|f^0\rangle{}$ state, 
originating from the increased energy of the $|f^2\rangle{}$ state. 
This suggests that $\Delta$ is essential for changing the property of the peak number, because $\Delta$ controls the energy of the initial and final $|f^2\rangle{}$ states. 
In fact, the most significant change between 
the parameters of La$_2$O$_3$ and 
CeO$_2$ is $\Delta$ (Table \ref{table1}). 
Thus, when we discuss the properties of the 
La$_2$O$_3$ and CeO$_2$ spectra, 
we can ignore the difference in parameters between 
La$_2$O$_3$ and CeO$_2$, excluding $\Delta$. 

\section{Recognition Model of Emulated Spectrum}
\label{recog_model}
In this section, we describe the two effective models for recognizing the emulated measurement data generated following the generative model $\mathcal{H}$. 
Hereinafter, we refer to these effective models as the recognition models. 
In this section, we introduce two recognition models referred to as the two-state Hamiltonian model and the three-state Hamiltonian model. 

\subsection{Two-state Hamiltonian model}
Kotani and Toyozawa\cite{Kotani74} proposed a two-state Hamiltonian model as an effective model of La$_2$O$_3$ XPS spectra. 
The two-state Hamiltonian model is defined using 
an effective Hamiltonian on two eigenstates 
of $4f$ electrons $|f^0\rangle{}$ and $|f^1\rangle{}$. 
The effective Hamiltonian is given as 
\begin{equation}
\begin{split}
H_2 &= \epsilon_L \sum_{\mu}   a^{\dagger}_{L\mu}a_{L\mu} + \epsilon^0_f \sum_{\mu} a^{\dagger}_{f\mu}a_{f\mu} + \epsilon_c a^{\dagger}_c a_c\\
&+ \frac{V}{\sqrt{N_f}} \sum_{\mu}  (a^{\dagger}_{L\mu}a_{f\mu} + a_{L\mu}a^{\dagger}_{f\mu}) -U_{fc} \sum_{\mu} a^{\dagger}_{f\mu}a_{f\mu}(1- a^{\dagger}_c a_c),
\end{split}
\end{equation}
where, as in the generative model Hamiltonian $\mathcal{H}$, $\epsilon_L$, $\epsilon^0_f$, and $\epsilon_c$ are the energies of the conducting electron of $4f$ rare-earth metals ($5d$, $6s$ electrons), the $4f$ electron, and the core electron, respectively. 
The index $\nu$ $(\nu=1 $--$ N_f, N_f=14)$ represents the quantum number of the spin and $f$-orbital. 
$V$ and $-U_{fc}$ are the energies of the hybridization interaction and the core-hole Coulomb potential for the $4f$ electrons, respectively. 
\par
We define the initial eigenstate of the minimum energy $E_G(\boldsymbol{\vartheta})$ as $|G\rangle{}$ and the final eigenstates of the two energy levels $E_j(\boldsymbol{\vartheta})$ $(j = 0,1)$  as $|F_j\rangle{}$ $(j = 0,1)$. 
In the two-state Hamiltonian model, 
the initial eigenstate $|G\rangle{}$ is set to be equal to 
the $4f$ electron eigenstate $|f^0\rangle{}$. 
By using Fermi's golden rule, 
convoluting the transition probability with the Lorentz function, 
and adding the Gaussian noise $\epsilon$, 
we obtained the spectral distribution
\begin{equation}
\begin{split}
\label{2state_model_intensity}
I_2(\omega;\boldsymbol{\theta}_2) = 
\sum_{j=0}^1 |\langle{}F_j|a_c|G\rangle{}|^2 \frac{\Gamma_{j}/\pi}{(\omega-(E_j(\boldsymbol{\theta}_2)-E_g(\boldsymbol{\theta}_2) - b))^2+\Gamma^2_j} + \epsilon,
\end{split}
\end{equation}
where, as in the generative model $\mathcal{H}$, we set the energy level of the initial and final states $|f^0\rangle{}$ to 0 as the standard energy level. 
Thus, the number of parameters of the two-state Hamiltonian model is five: $\Delta^{\prime} (= \epsilon^0_f - \epsilon_L - U_{fc}), V , \Gamma_1,\Gamma_2$, and $b$. 
The energy shift parameter $b$ is added to the model to compensate for the difference in the standard energy level between the models. 
The energy shift parameter is also added to the model in the conventional analysis of the XPS spectra. 
Because the two-state Hamiltonian model is uniquely defined by the effective Hamiltonian $H_2$, we label this model as $H_2$. 
\par
It was reported that the XPS spectra of La$_2$O$_3$ can be reproduced by the two-state Hamiltonian model $H_2$\cite{Kotani74}. 
As we mentioned, 
the emulated measurement data of the La$_2$O$_3$ XPS spectrum is generated on the basis of the generative model $\mathcal{H}$, which is more complex than the two-state Hamiltonian model $H_2$. 
One purpose of this study is to confirm that this physical knowledge is reproduced by the proposed method. 

\subsection{Three-state Hamiltonian model}
Kotani et al.\cite{Kotani85} proposed a three-state Hamiltonian model as an effective model of CeO$_2$ XPS spectra. 
The effective Hamiltonian of the three-state Hamiltonian model 
is defined as
\begin{equation}
\begin{split}
 H_3 &= \epsilon_L \sum_{\nu}  a^{\dagger}_{L\nu}a_{L\nu} + \epsilon^0_f \sum_{\nu} a^{\dagger}_{f\nu}a_{f\nu} + \epsilon_c a^{\dagger}_c a_c\\
&+ \frac{V}{\sqrt{N_f}} \sum_{\nu}  (a^{\dagger}_{f\nu}a_{f\nu} + a_{f\nu}a^{\dagger}_{f\nu}) + U_{ff} \sum_{\nu\rangle{}\nu'} a^{\dagger}_{f\nu}a_{f\nu}a^{\dagger}_{f\nu'}a_{f\nu'}\\
&-U_{fc} \sum_{\nu} a^{\dagger}_{f\nu}a_{f\nu}(1- a^{\dagger}_c a_c).
\end{split}
\end{equation}
Then, as in the previously described model, 
the XPS spectral model is derived as
\begin{equation}
\begin{split}
I_3(\omega;\boldsymbol{\theta}_3) =
 \sum_{j=0}^2 |\langle{}F_j|a_c|G\rangle{}|^2 \frac{\Gamma_{j}/\pi}{(\omega-(E_j(\boldsymbol{\theta}_3)-E_G(\boldsymbol{\theta}_3) - b))^2+\Gamma^2_j} + \epsilon,
\end{split}
\end{equation}
where $|G\rangle{}$ is the initial eigenstate of the minimum energy $E_G(\boldsymbol{\vartheta})$, and $E_j(\boldsymbol{\vartheta})$ $(j = 0,1,2)$ as $|F_j\rangle{}$ $(j = 0,1,2)$ are the final eigenstates of the three energy levels. 
We set the energy level of the initial and final states $|f^0\rangle{}$ to 0 as the standard energy level. 
Thus, the number of parameters of the three-state Hamiltonian model is eight: $\Delta (= \epsilon^0_f - \epsilon_L), V , U_{ff}, U_{fc}, \Gamma_1, \Gamma_2, \Gamma_3$, and $b$. 
The three-state Hamiltonian model and the generative model are similar except for $b$ and the degree of freedom of $\Gamma$. 
\par
As we mentioned earlier, 
it was reported that the XPS spectra of CeO$_2$ can be reproduced by the three-state Hamiltonian model $H_3$\cite{Kotani85}. 
Because the three-state Hamiltonian model is uniquely defined by the effective Hamiltonian $H_3$, we label this model as $H_3$. 

\section{Method}
\subsection{Bayesian model selection}
We evaluate the recognition models $H_2$ and $H_3$ 
in terms of their capability to represent the spectrum data $\boldsymbol{D}=\{\boldsymbol{w},\boldsymbol{\mathcal{I}}\}=\{(w_1,w_2,\cdots w_N),(\mathcal{I}(w_1;\boldsymbol{\vartheta}),\mathcal{I}(w_2;\boldsymbol{\vartheta}),\cdots \mathcal{I}(w_N;\boldsymbol{\vartheta}))\}$ generated by 
the generative model $\mathcal{H}$. 
The likelihood of the recognition model $H_k(k=2,3)$ for the dataset $\boldsymbol{D}$ is defined as
\begin{equation}
{\rm P}(H_k|\boldsymbol{D}) = \frac{{\rm P}(\boldsymbol{D}|H_k){\rm P}(H_k)}{{\rm P}(\boldsymbol{D})}  \propto  {\rm P}(\boldsymbol{D}|H_k){\rm P}(H_k),
\end{equation}
where ${\rm P}(\boldsymbol{D})$ is a normalization constant. 
In this study, we assume that there is no prior knowledge 
about the likelihood of the model. Thus, we set the prior probability ${\rm P}(H_k)$ 
as a uniform distribution; in this study, it is equal to $\frac{1}{2}$. 
We also assume that $\boldsymbol{w}$ in the dataset 
$\boldsymbol{D}=\{\boldsymbol{w},\boldsymbol{\mathcal{I}}\}$ 
is given deterministically, that is, non-probabilistically. 
Then, the likelihood of the model is transformed as 
\begin{equation}
\begin{split}
{\rm P}(H_k|\boldsymbol{D}) &\propto {\rm P}(\boldsymbol{D}|H_k) = {\rm P}(\boldsymbol{\mathcal{I}}|H_k)\\
&= \int_{-\infty}^{\infty} {\rm P}({\boldsymbol{\mathcal{I}}}|\boldsymbol{\theta}_k,H_k) {\rm P}(\boldsymbol{\theta}_k|H_k) d\boldsymbol{\theta}_k, 
\end{split}
\label{like}
\end{equation}
where $\boldsymbol{\theta}_k$ is the parameter set of the recognition model 
$H_k$ described in Sect. \ref{recog_model}. 
\par
The conditional probability ${\rm P}({\boldsymbol{\mathcal{I}}}|\boldsymbol{\theta}_k,H_k)$ of 
Eq. (\ref{like}) is a stochastic generative model of the recognition model 
$H_k$. 
When the additive noise $\epsilon$ of the XPS spectra 
is given as an independent and identically distributed 
Gaussian with average 0 and 
standard deviation $\sigma_{noise}$, 
\begin{equation}
\begin{split}
\label{noise_model}
&{\rm P}({\boldsymbol{\mathcal{I}}}|\boldsymbol{\theta}_k,H_k) = \prod_{i=1}^N {\rm P}(\mathcal{I}(w_i)|\boldsymbol{\theta}_k,H_k)\\
&=\Bigl(\frac{1}{2\pi\sigma_{noise}^2}\Bigr)^{N/2} \prod_{i=1}^N \exp\left[ -\frac{1}{2\sigma_{noise}^2}(\mathcal{I}(w_i;\boldsymbol{\vartheta}) - I_k(w_i;\boldsymbol{\theta}_k))^2 \right] \\
&=\Bigl(\frac{1}{2\pi\sigma_{noise}^2}\Bigr)^{N/2} \exp\Bigl\{ -\sum_{i=1}^N \left[\frac{1}{2\sigma_{noise}^2}(\mathcal{I}(w_i;\boldsymbol{\vartheta}) - I_k(w_i;\boldsymbol{\theta}_k))^2\right] \Bigr\}.
\end{split}
\end{equation}
The probability $ {\rm P}(\boldsymbol{\theta}_k|H_k)$ in 
Eq. (\ref{like}) simulates the prior knowledge about the model parameters 
$\boldsymbol{\theta}_k$ as a probability distribution. 
By substituting Eq. (\ref{noise_model}) into Eq. (\ref{like}), 
we obtain the following:
\begin{equation}
\begin{split}
&{\rm P}(\boldsymbol{\mathcal{I}}|H_k) = \Bigl(\frac{1}{2\pi\sigma_{noise}^2}\Bigr)^{N/2} \\
& \times \int_{-\infty}^{\infty} \exp\biggl\{ -\frac{1}{2\sigma_{noise}^2}\sum_{i=1}^N \left[\mathcal{I}(w_i;\boldsymbol{\vartheta}) - I_k(w_i;\boldsymbol{\theta}_k)\right]^2\biggr\}{\rm P}(\boldsymbol{\theta}_k|H_k)  d\boldsymbol{\theta}_k\\
&=\Bigl(\frac{1}{2\pi\sigma_{noise}^2}\Bigr)^{N/2}  \int_{-\infty}^{\infty} \exp\left[ -NE(\boldsymbol{\theta}_k)  \right]{\rm P}(\boldsymbol{\theta}_k|H_k) d\boldsymbol{\theta}_k,
\end{split}
\label{like2}
\end{equation}
where 
\begin{equation}
\label{energy}
E(\boldsymbol{\theta}_k) = \frac{1}{2N\sigma_{noise}^2}\sum_{i=1}^N  \left[ \mathcal{I}(w_i;\boldsymbol{\vartheta}) - I_k(w_i;\boldsymbol{\theta}_k)\right] ^2.
\end{equation}
The probability ${\rm P}(\boldsymbol{\mathcal{I}}|H_k)$ is often referred to as the marginal likelihood and is proportionally related to the likelihood of the recognition model $H_k$. 
The negative log-likelihood,
\begin{equation}
F(H_k)=-\log {\rm P}(\boldsymbol{\mathcal{I}}|H_k),
\label{FE}
\end{equation}
is often referred to as the Bayesian free energy ($FE$). 
The effective model $H_k$ with the smallest $FE$ value
represents the best model.

\subsection{Exchange Monte Carlo method}
To obtain the value of $FE$, we need to execute the integration in Eq. (\ref{like2}). 
However, it is difficult to analytically execute the integration 
owing to the complicated relationship between $\boldsymbol{\mathcal{I}}$ 
and $\boldsymbol{\theta}_k$. 
We overcame this difficulty by numerical integration using the exchange Monte Carlo method\cite{Hukushima96}. 
\par
Markov chain Monte Carlo (MCMC) methods\cite{metropolis53} are efficient for sampling from a probability distribution in a high-dimensional space, such as $\boldsymbol{\theta}_k$. 
To apply an MCMC method to execute an integration, 
we need to transform the integration to a mean value calculation. 
When applying an MCMC method to the calculation of $FE$, 
we introduce an auxiliary variable $\beta$ 
and transform Eq. (\ref{FE}) into 
\begin{equation}
\begin{split}
\label{integ_beta}
&F(H_k)=-\log\int_{-\infty}^{\infty} \exp\left[-NE(\boldsymbol{\theta}_k)\right]{\rm P}(\boldsymbol{\theta}_k|H_k) d\boldsymbol{\theta}_k\\
&= \int^1_0 \frac{\partial}{\partial \beta}\left\{ -\log\left[ \int_{-\infty}^{\infty} \exp(-\beta N E(\boldsymbol{\theta}_k)){\rm P}(\boldsymbol{\theta}_k|H_k)d\boldsymbol{\theta}_k\right]\right\}d\beta \\
&=  \int^1_0 \int_{-\infty}^{\infty} NE(\boldsymbol{\theta}_k){\rm P}(\boldsymbol{\theta}_k|\boldsymbol{\mathcal{I}},\beta) d\boldsymbol{\theta}_k d\beta\\
&= \int^1_0 <NE(\boldsymbol{\theta}_k)>_{{\rm P}(\boldsymbol{\theta}_k|\boldsymbol{\mathcal{I}},\beta)} d\beta,
\end{split}
\end{equation}
where $<\cdot>$ represents an average and 
\begin{equation}
\label{posterior_prob}
{\rm  P}(\boldsymbol{\theta}_k|\boldsymbol{\mathcal{I}},\beta) = \frac{\exp[-\beta N E(\boldsymbol{\theta}_k)]{\rm  P}(\boldsymbol{\theta}_k|H_k)}{\int_{-\infty}^{\infty}  \exp[-\beta N E(\boldsymbol{\theta}_k)]{\rm  P}(\boldsymbol{\theta}_k|H_k)d\boldsymbol{\theta}_k}.
\end{equation}
When $N E(\boldsymbol{\theta}_k)$ is regarded as energy, 
this equation suggests that ${\rm  P}(\boldsymbol{\theta}_k|\boldsymbol{\mathcal{I}},\beta)$ corresponds to the Boltzmann distribution in statistical physics. 
In the same way, the auxiliary variable $\beta$ corresponds to the inverse temperature in statistical physics. 
Equation (\ref{integ_beta}) is approximated to a quadrature by parts, 
\begin{equation}
F(H_k) \simeq \sum_{l=0}^L <NE(\boldsymbol{\theta}_k)>_{{\rm P}(\boldsymbol{\theta}_k|\boldsymbol{\mathcal{I}},\beta_l)} \Delta \beta_l,
\end{equation}
where $\beta_l$ is given as a sequence of inverse temperatures $0=\beta_1<\beta_1<\cdots<\beta_L=1$ obtained by dividing $\beta=0$ to $\beta=1$ into $L$ pieces, 
and $<NE(\boldsymbol{\theta}_k)>_{{\rm P}(\boldsymbol{\theta}_k|\boldsymbol{\mathcal{I}},\beta_l)}$ is obtained by 
$\beta_l$-independent MCMC sampling. 
However, the MCMC sampling is often trapped at local minima. 
\par
The exchange Monte Carlo method (EMC) is an algorithm of an MCMC method used to avoid local trapping at minima. 
This method simulates multiple samplings from multiple densities with different inverse temperatures $\{\theta_l\}_{l=1}^L$. 
The EMC takes samples from the joint density 
  \begin{equation}
{\rm  P}(\boldsymbol{\theta}_k^1,\boldsymbol{\theta}_k^2\cdots \boldsymbol{\theta}_k^L|\boldsymbol{\mathcal{I}})=\prod_{l=1}^L{\rm  P}(\boldsymbol{\theta}_k^l|\boldsymbol{\mathcal{I}},\beta_l),
\end{equation}
where the probability density ${\rm  P}(\boldsymbol{\theta}_k^l|\boldsymbol{\mathcal{I}},\beta_l)$ is defined in Eq. (\ref{posterior_prob}). 
The EMC algorithm is based on the following updates, in which the joint density ${\rm  P}(\boldsymbol{\theta}_k^1,\boldsymbol{\theta}_k^2\cdots \boldsymbol{\theta}_k^L|\boldsymbol{\mathcal{I}})$ is invariant.
 \\

\begin{itemize}
\item[{\bf 1}]{\bf Sampling from each density ${\rm  P}(\boldsymbol{\theta}_k^l|\boldsymbol{\mathcal{I}},\beta_l)$}\par
Sampling from ${\rm  P}(\boldsymbol{\theta}_k^l|\boldsymbol{\mathcal{I}},\beta_l)$ by a conventional MCMC method, 
such as \par the Metropolis\textendash Hastings algorithm\cite{hastings70}. 

 \item[{\bf 2}] {\bf Exchange process between two densities corresponding to adjacent inverse temperatures}\par
 The exchanges between the configurations $\boldsymbol{\theta}_k^l$ and 
 $\boldsymbol{\theta}_k^{l+1}$ correspond to adjacent inverse temperatures following the probability $R=\min(1,r)$, where\par
  
\begin{equation}
\begin{split}
\;\;\;\;\; r &= \frac{{\rm  P}(\boldsymbol{\theta}_k^1,\cdots, \boldsymbol{\theta}_k^{l+1}, \boldsymbol{\theta}_k^l,\cdots,\boldsymbol{\theta}_k^L|\boldsymbol{\mathcal{I}})}{{\rm  P}(\boldsymbol{\theta}_k^1,\cdots, \boldsymbol{\theta}_k^l, \boldsymbol{\theta}_k^{l+1},\cdots,\boldsymbol{\theta}_k^L|\boldsymbol{\mathcal{I}})}\\
&=\frac{{\rm  P}(\boldsymbol{\theta}_k^{l+1}|\boldsymbol{\mathcal{I}},\beta_l){\rm  P}(\boldsymbol{\theta}_k^l|\boldsymbol{\mathcal{I}},\beta_{l+1})}{{\rm  P}(\boldsymbol{\theta}_k^l|\boldsymbol{\mathcal{I}},\beta_l){\rm  P}(\boldsymbol{\theta}_k^{l+1}|\boldsymbol{\mathcal{I}},\beta_{l+1})}\\
&=\exp\left\{N[\beta_{l+1} - \beta_{l}][E(\boldsymbol{\theta}_k^{l+1}) - E(\boldsymbol{\theta}_k^l)]\right\}.
\end{split}
\end{equation}

\end{itemize}
Sampling from a distribution with a smaller $\beta$ 
corresponds to sampling from a distribution with a larger intensity of noise; thus, the distribution tends not to have a local minimum. 
On the other hand, sampling from a distribution with 
a larger $\beta$ corresponds to sampling from a distribution with 
local minima. 
Hence, sampling from the joint density ${\rm  P}(\boldsymbol{\theta}_k^1,\boldsymbol{\theta}_k^2\cdots \boldsymbol{\theta}_k^L|\boldsymbol{\mathcal{I}})$ overcomes the 
local minimum and enables the fast convergence of sampling. 
\par
Using the sampling result of the $\beta=1$ state, 
we obtained a posterior density of the parameter ${\rm P}(\boldsymbol{\theta}_k|\boldsymbol{\mathcal{I}})$ [Eq. (\ref{posterior_prob})]. 
From the posterior density of $\boldsymbol{\theta}_k$, 
we can estimate the model parameters $\boldsymbol{\theta}_k$ 
of $H_k$ and the related information, such as estimation accuracy. 

\section{Results}

\begin{figure*}[ht]
  \begin{center}
    \includegraphics[clip,width=15.0cm]{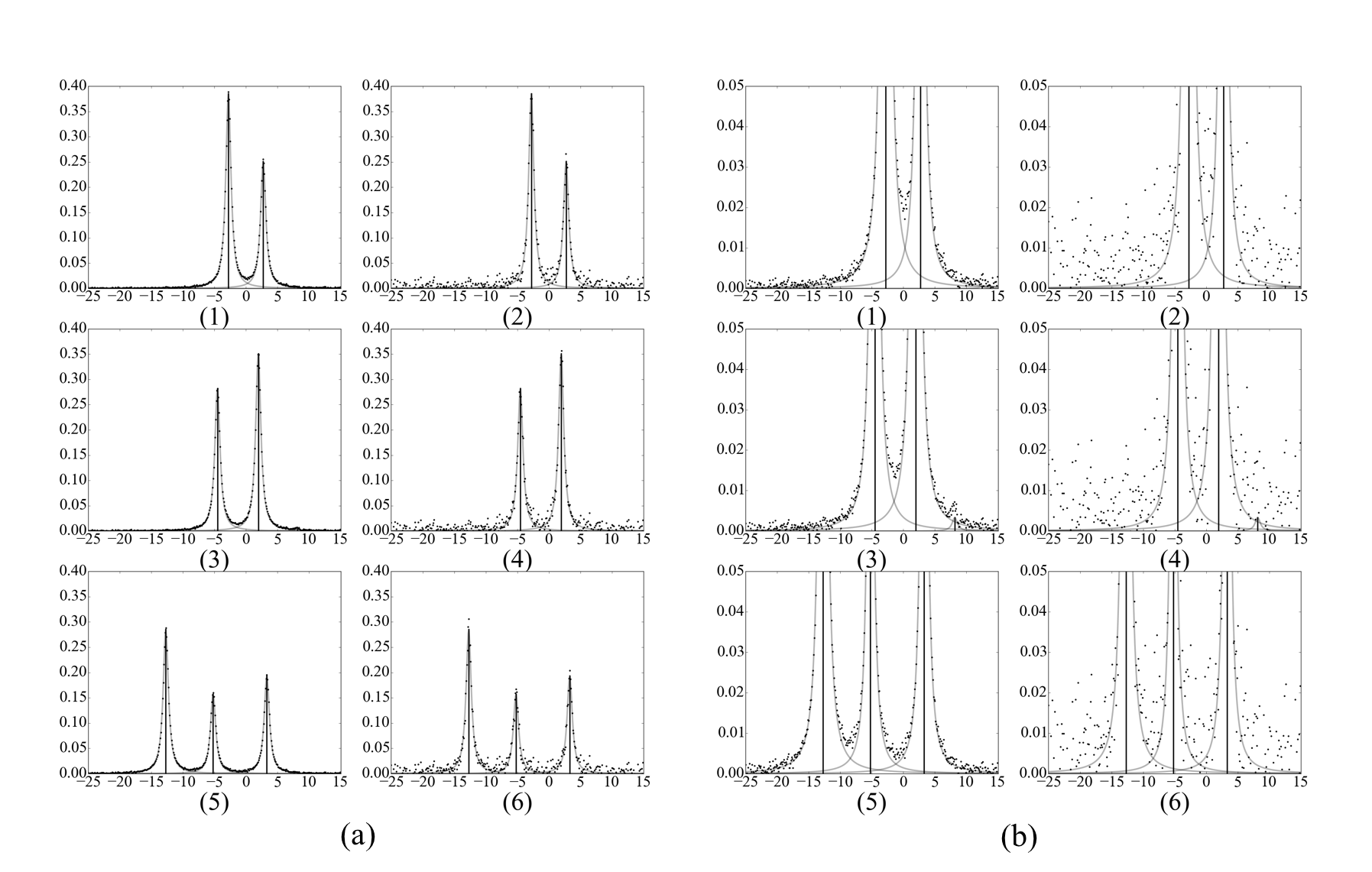}
    \caption{Graphs of spectra generated by the generative model $\mathcal{H}$ with parameters $V=0.76, U_{ff}=10.5, U_{fc}=12.5$, and $\Gamma = 0.7$. Gray dots represent the observed data, the curved gray line represents the true spectral density, and the black line represents the peak position and peak intensity. For $\Delta$ and $\sigma_{noise}$, the following values were set for each spectrum:  (a)-(1): [$\Delta$，$\sigma_{noise}$] = [12.5, 0.001], (a)-(2): [12.5, 0.01], (a)-(3): [10.08, 0.001], (a)-(4): [10.08, 0.01], (a)-(5): [1.6, 0.001], (a)-(6): [1.6, 0.01] and (b): Enlarged plots of spectra in (a).}
    \label{fig_spectrals}
  \end{center}
\end{figure*}

\begin{table*}[t]
  \caption{Properties of spectrum structure. $E_j - E_g$ corresponds to the peak position, and $\left|\langle{}F_j|a_c|G\rangle{}\right|^2\frac{\Gamma/\pi}{E_j - E_g - (E_j - E_g) + \Gamma^2}=\frac{|\langle{}F_j|a_c|G\rangle{}|^2}{\Gamma \pi}$ 
  corresponds to the peak intensity. The number of peaks is defined as the number of peaks whose intensity is larger than the noise intensity $\sigma_{noise}$.}
\begin{center}
  \begin{tabular}{c||c|c|c|c|c|c|c} \hline
   $\Delta$  & Number of peaks & $E_0 - E_g$ & $E_1 - E_g$ & $E_2 - E_g$ & $\frac{|\langle{}F_0|a_c|G\rangle{}|^2}{\Gamma \pi}$ & $\frac{|\langle{}F_1|a_c|G\rangle{}|^2}{\Gamma \pi}$ & $\frac{|\langle{}F_2|a_c|G\rangle{}|^2}{\Gamma \pi}$\\ \hline \hline
12.5 & 2 & -2.8 & 2.72 & 12.48 & 0.3850 & 0.2512 & 0.0005\\ \hline 
10.08 & 2-3 & -4.53 & 1.94 & 8.15 & 0.2824 & 0.3509 & 0.0033\\ \hline 
7.66 & 3 & -6.67 & 1.25 & 4.3746 & 0.2330 & 0.3581 & 0.0456\\ \hline 
5.23 & 3 & -9.03 & 2.74 & -0.92 & 0.2298 & 0.2414 & 0.1654\\ \hline 
2.81 & 3 & 2.91 & -11.5 & -3.93 & 0.2372 & 0.2623 & 0.1372\\ \hline 
1.6 & 3 & 3.29 & -5.22 & -12.71 & 0.1937 & 0.1575 & 0.2854\\ \hline 
  \end{tabular}
  \label{table_spectral}
\end{center}
\end{table*}

\begin{table*}[t]
  \caption{Range of uniform prior densities of parameters.}
\begin{center}
  \begin{tabular}{c||c|c|c|c|c|c|c} \hline
     & $\Delta$ & $\Delta^{\prime}$ &$V$ & $U_{ff}$ & $U_{fc}$ & $\Gamma$ & b\\ \hline \hline
    Min & 0.0  &-20.0& 0.0 & 0.0 & 0.0 & 0.01 & -5.0    \\ \hline 
    Max & 20.0 &20.0& 4.0 & 20.0 & 20.0 & 1.0 & 5.0  \\ \hline  
  \end{tabular}
  \label{table2}
\end{center}
\end{table*}

\begin{figure*}[t]
  \begin{center}
    \includegraphics[clip,width=14.0cm]{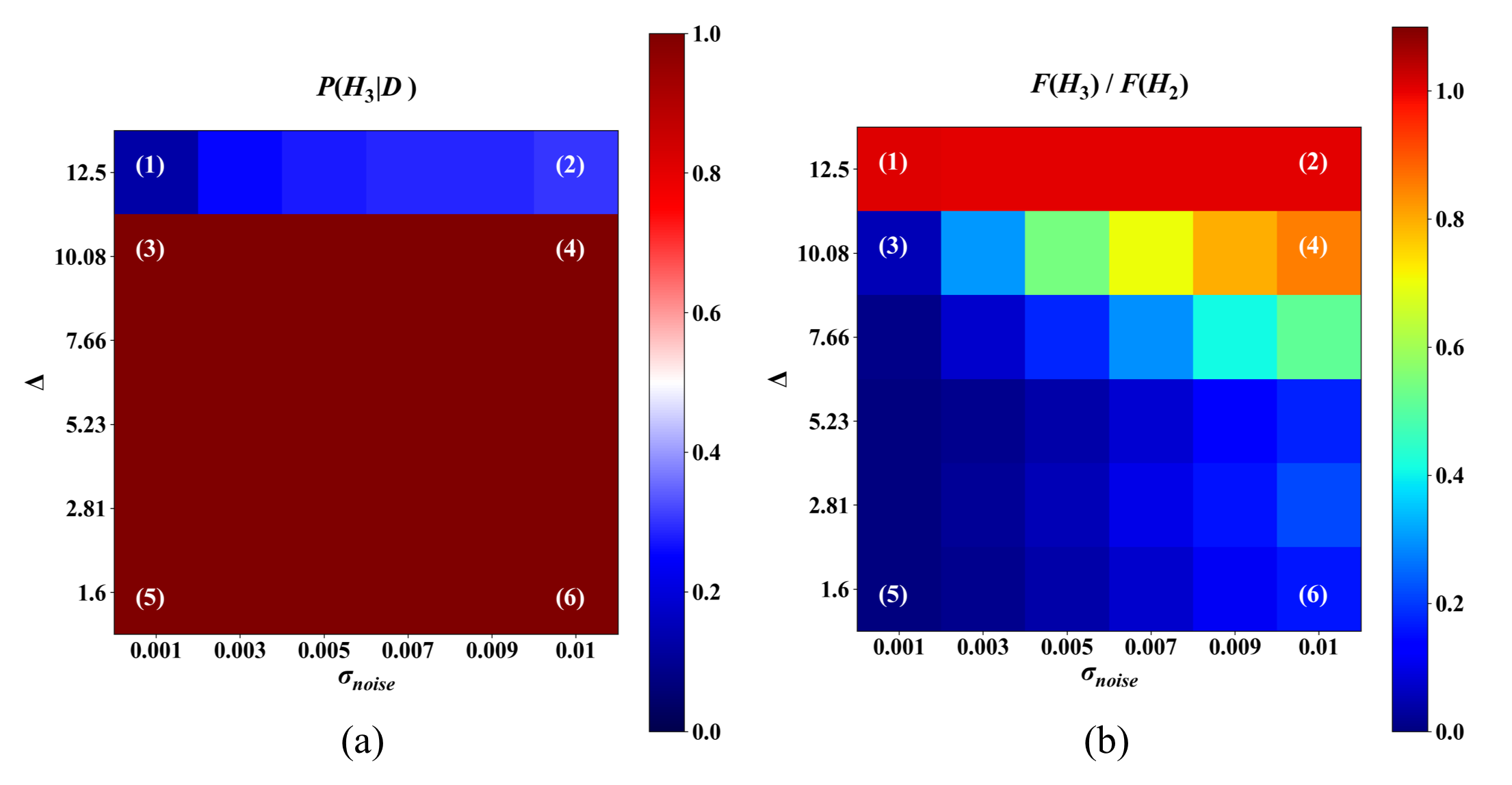}
    \caption{(a) Likelihood of the $H_3$ Hamiltonian model. Colored cells represents the value of ${\rm P}(H_3|\boldsymbol{D})$. (b) Ratio of the Bayesian free energy of the $H_3$ Hamiltonian model to that of $H_2$. Colored cells represent the value of $F(H_3)/F(H_2)$. }
    \label{fig_heatmap}
  \end{center}
\end{figure*}

\begin{figure*}[t]
  \begin{center}
    \includegraphics[clip,width=9.5cm]{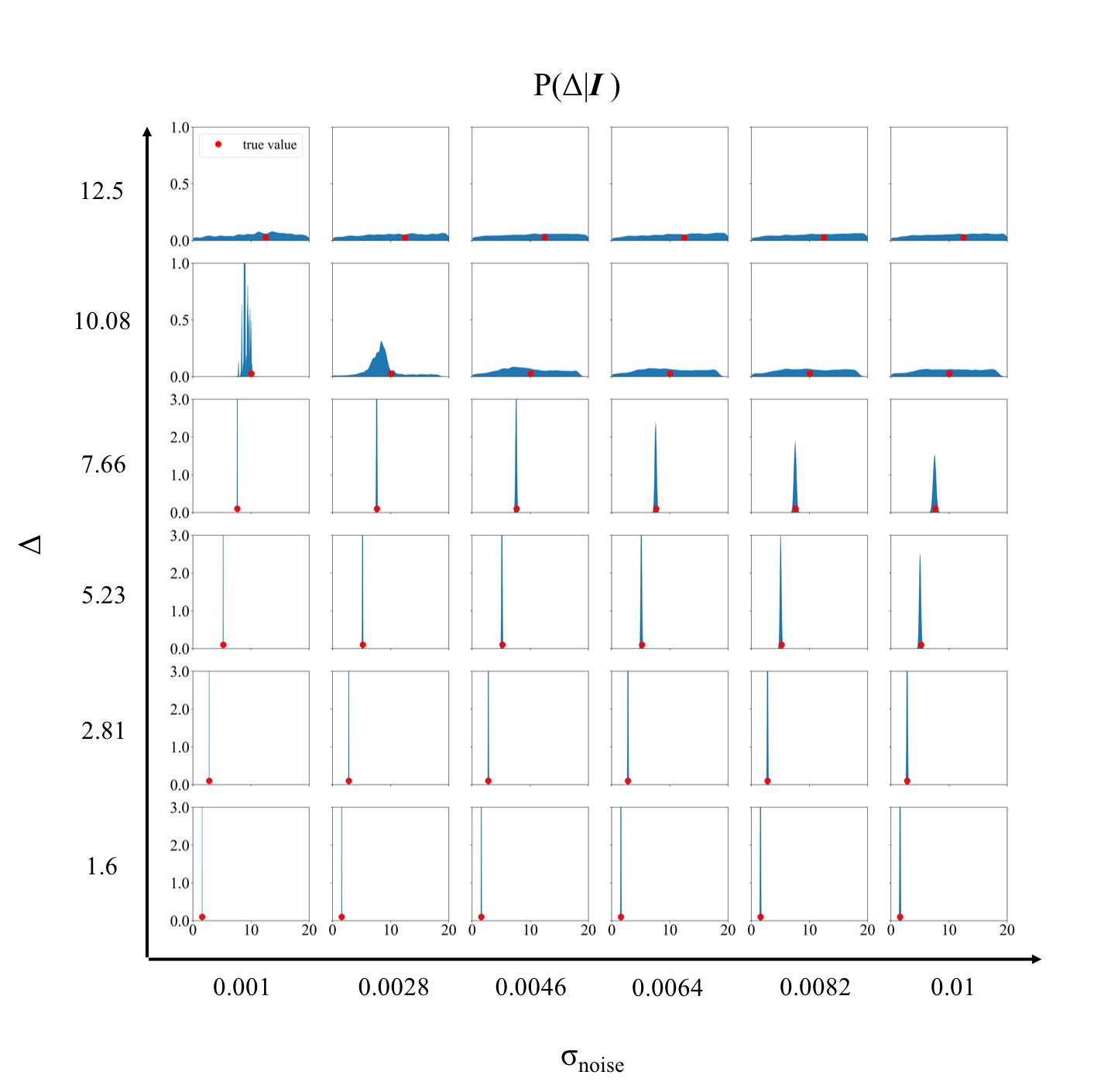}
    \caption{Marginalized posterior densities of $\Delta$ arranged in the same order as in the heat map of Fig. \ref{fig_heatmap}. The horizontal axis of each graph is $\Delta$ and the vertical axis is $P(\Delta|\boldsymbol{I})$. The range of horizontal axis of all graphs is from 0.0 to 20.0, and the range of vertical axis of the graphs with $\Delta$ $<$ 10.08 is from 0.0 to 3.0 and the graphs with $\Delta$ $\geq$ 10.08 is from 0.0 to 1.0. The red circle described in the graph indicates the true parameter, which we used to generate the emulated measurement data.}
    \label{fig_post_delta}
  \end{center}

  \begin{center}
    \includegraphics[clip,width=9.5cm]{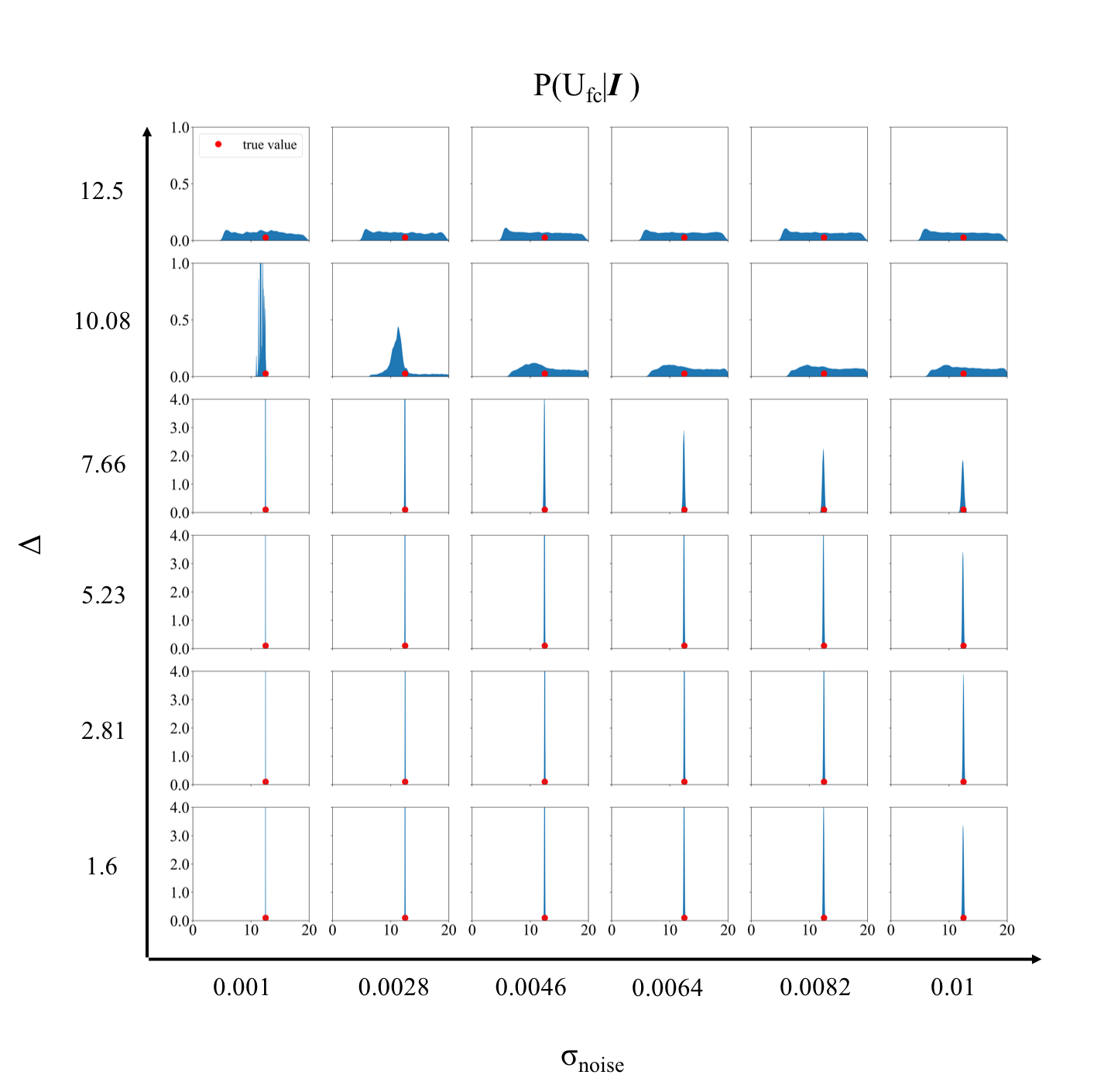}
    \caption{Marginalized posterior densities of $U_{fc}$ arranged in the same order as in the heat map of Fig. \ref{fig_heatmap}. The horizontal axis of each graph is $U_{fc}$ and the vertical axis is $P(U_{fc}|\boldsymbol{I})$. The range of horizontal axis of all graphs is from 0.0 to 20.0, and the range of vertical axis of the graphs with $\Delta$ $<$ 10.08 is from 0.0 to 4.0 and the graphs with $\Delta$ $\geq$ 10.08 is from 0.0 to 1.0. The red circle described in the graph indicates the true parameter, which we used to generate the emulated measurement data.}
    \label{fig_post_ufc}
  \end{center}
\end{figure*}

\begin{figure*}[t]
  \begin{center}
    \includegraphics[clip,width=9.5cm]{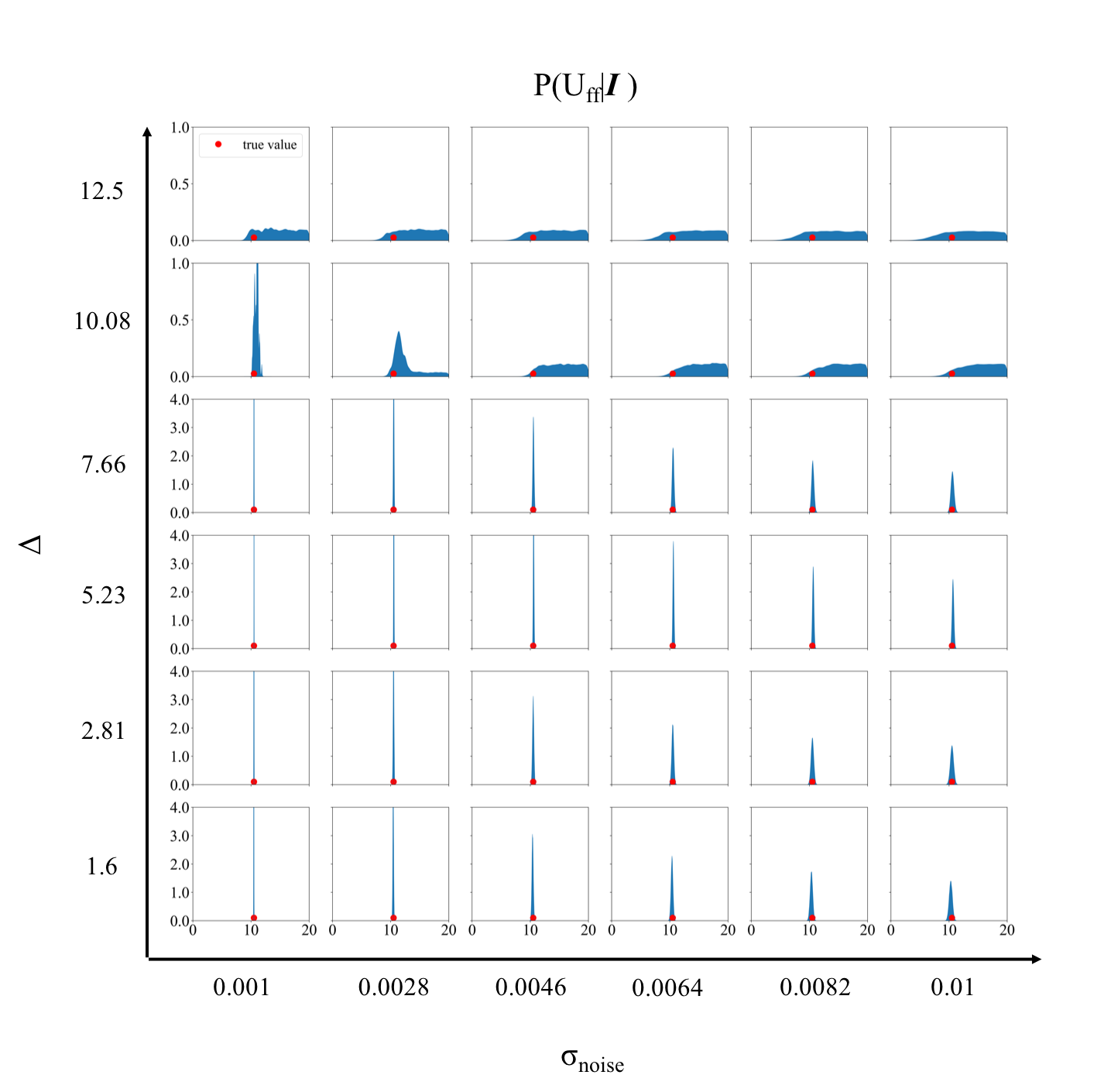}
    \caption{Marginalized posterior densities of $U_{ff}$ arranged in the same order as in the heat map of Fig. \ref{fig_heatmap}. The horizontal axis of each graph is $U_{ff}$ and the vertical axis is $P(U_{ff}|\boldsymbol{I})$. The range of horizontal axis of all graphs is from 0.0 to 20.0, and the range of vertical axis of the graphs with $\Delta$ $<$ 10.08 is from 0.0 to 4.0 and the graphs with $\Delta$ $\geq$ 10.08 is from 0.0 to 1.0. The red circle described in the graph indicates the true parameter, which we used to generate the emulated measurement data.}
    \label{fig_post_uff}
  \end{center}

  \begin{center}
    \includegraphics[clip,width=9.5cm]{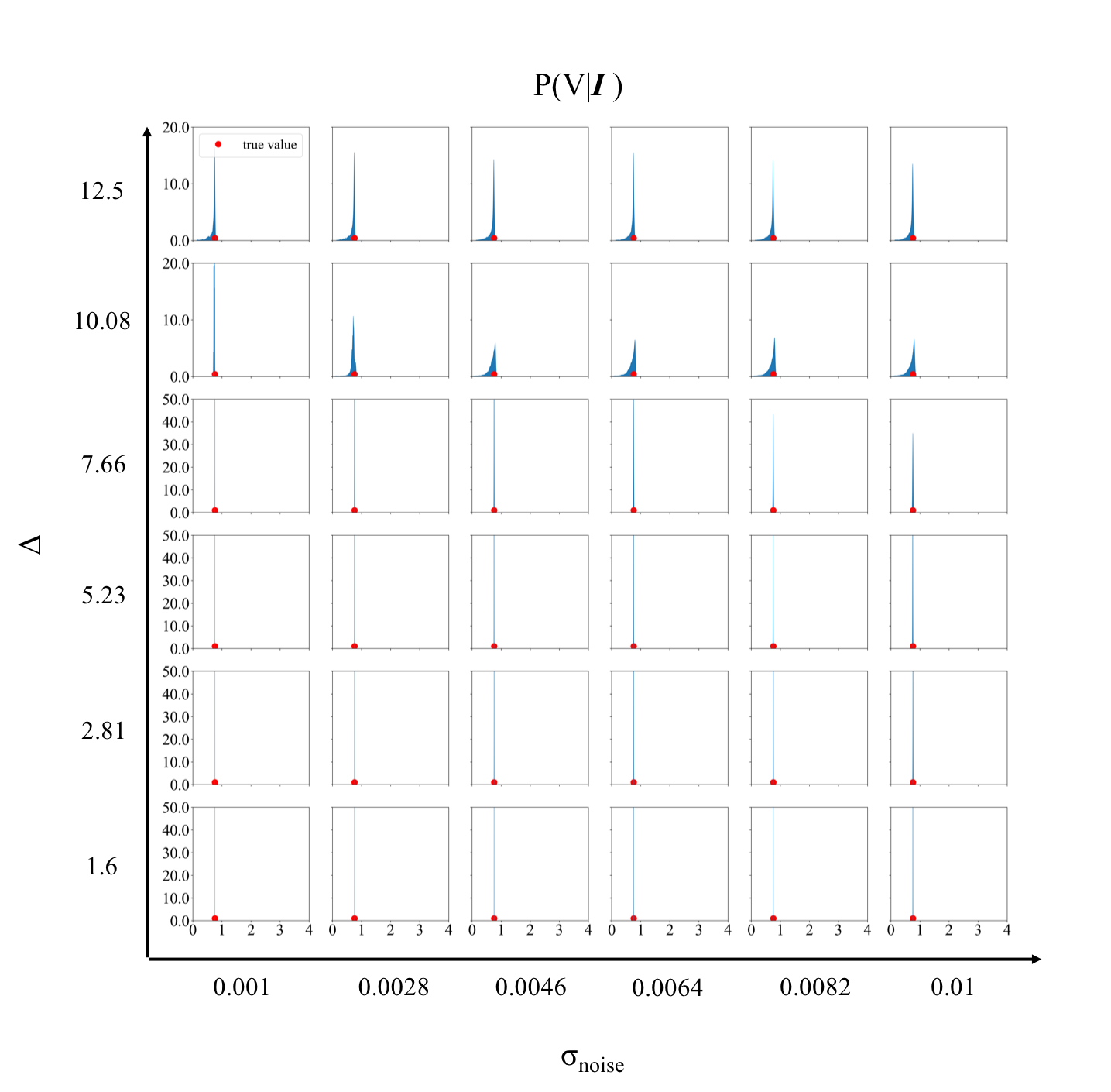}
    \caption{Marginalized posterior densities of $V$ arranged in the same order as in the heat map of Fig. \ref{fig_heatmap}. The horizontal axis of each graph is $V$ and the vertical axis is $P(V|\boldsymbol{I})$. The range of horizontal axis of all graphs is from 0.0 to 4.0, and the range of vertical axis of the graphs with $\Delta$ $<$ 10.08 is from 0.0 to 50.0 and the graphs with $\Delta$ $\geq$ 10.08 is from 0.0 to 20.0. The red circle described in the graph indicates the true parameter, which we used to generate the emulated measurement data.}
    \label{fig_post_v}
  \end{center}
\end{figure*}

\begin{figure*}[t]
  \begin{center}
    \includegraphics[clip,width=9.5cm]{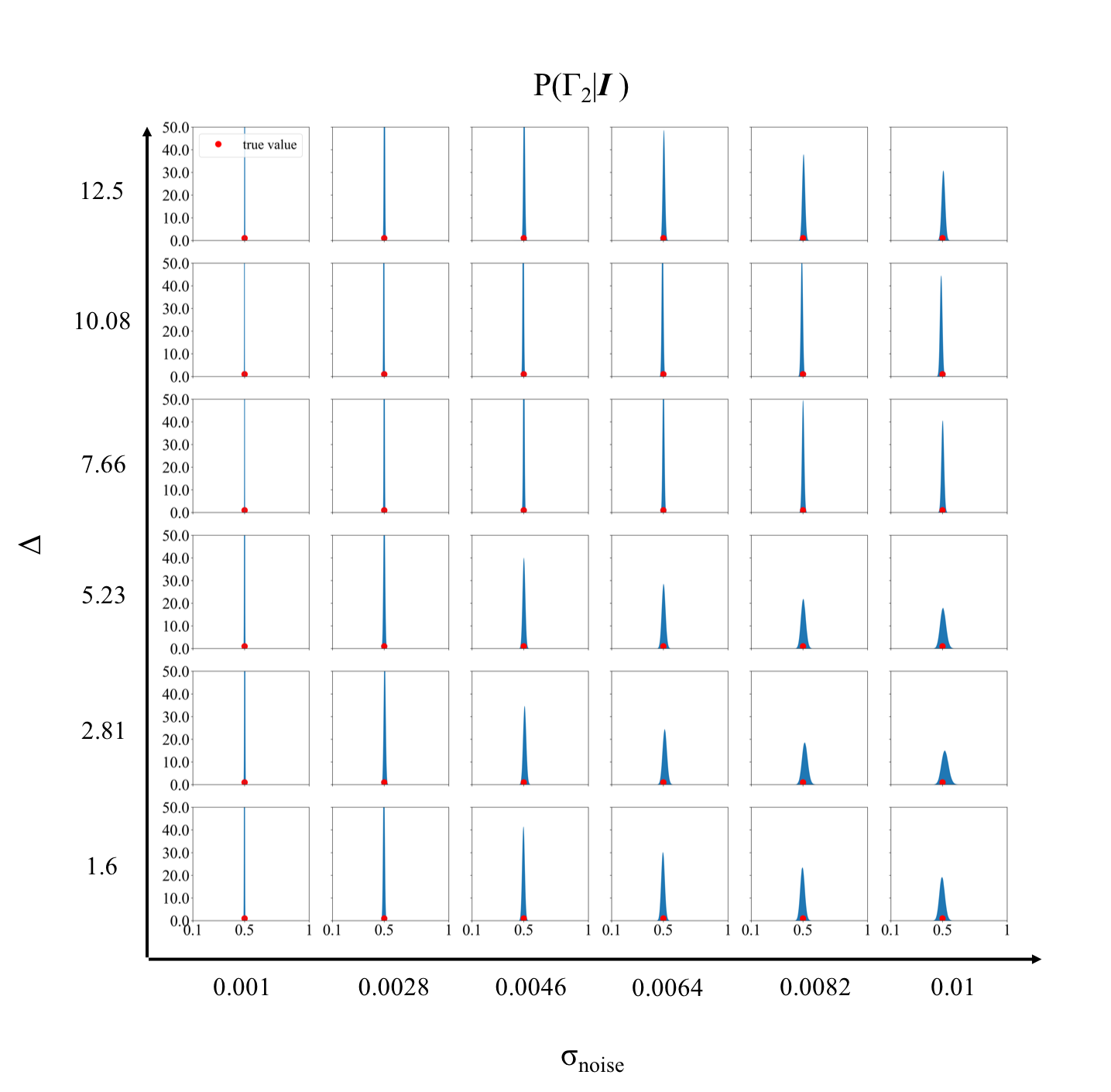}
    \caption{Marginalized posterior densities of $\Gamma_2$ arranged in the same order as in the heat map of Fig. \ref{fig_heatmap}. The horizontal axis of each graph is $\Gamma_2$ and the vertical axis is $P(\Gamma_2|\boldsymbol{I})$. The range of horizontal axis of all graphs is from 0.01 to 1.0, and the range of vertical axis of all graphs is from 0.0 to 50.0. The red circle described in the graph indicates the true parameter, which we used to generate the emulated measurement data.}
    \label{fig_post_gamma2}
  \end{center}

  \begin{center}
    \includegraphics[clip,width=9.5cm]{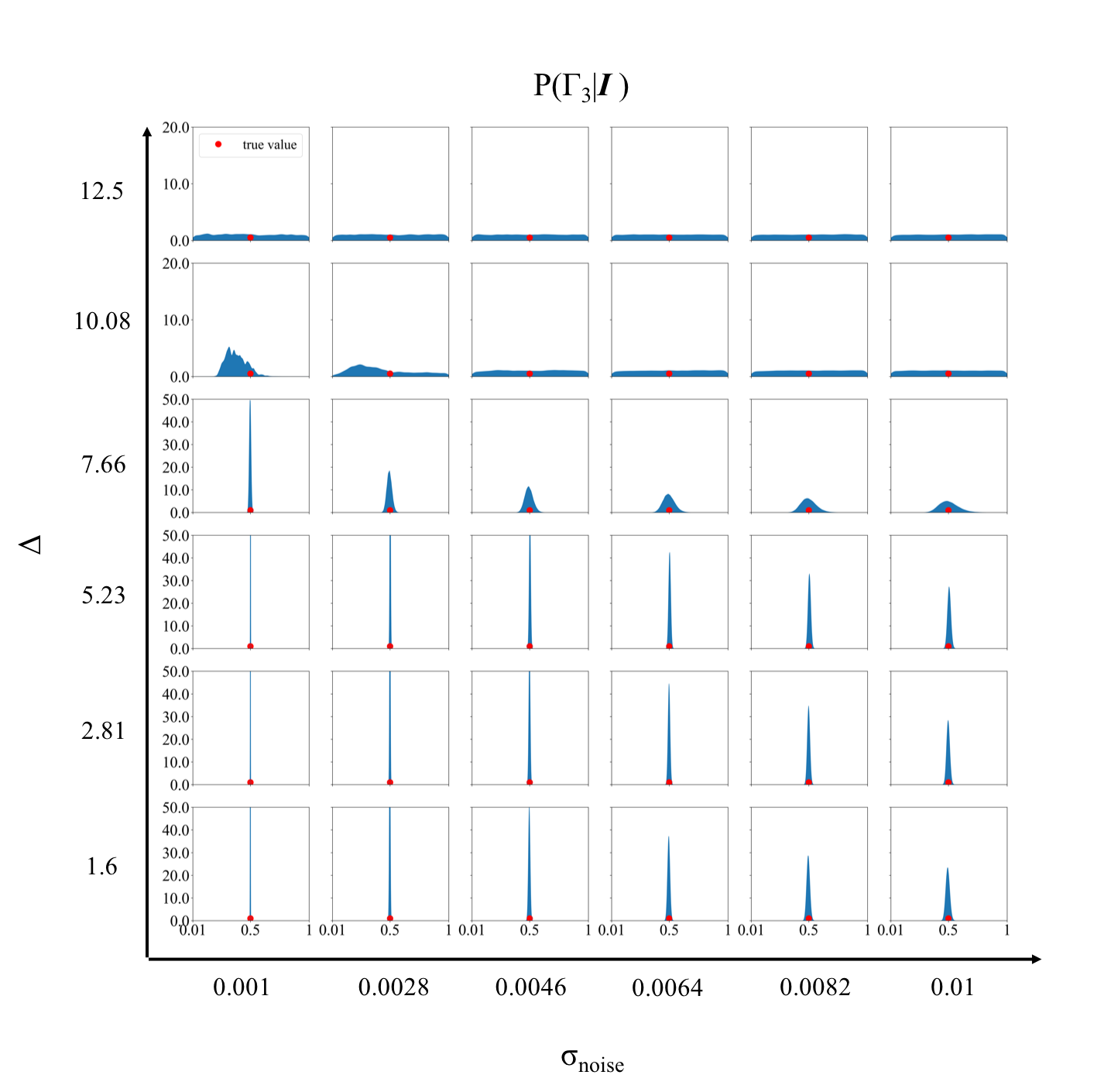}
    \caption{Marginalized posterior densities of $\Gamma_3$ arranged in the same order as in the heat map of Fig. \ref{fig_heatmap}. The horizontal axis of each graph is $\Gamma_3$ and the vertical axis is $P(\Gamma_3|\boldsymbol{I})$. The range of horizontal axis of all graphs is from 0.01 to 1.0, and the range of vertical axis of the graphs with $\Delta$ $<$ 10.08 is from 0.0 to 50.0 and the graphs with $\Delta$ $\geq$ 10.08 is from 0.0 to 20.0. The red circle described in the graph indicates the true parameter, which we used to generate the emulated measurement data.}
    \label{fig_post_gamma3}
  \end{center}
\end{figure*}

We applied our proposed method to the emulated measurement data and estimated the likelihood of the effective models $H_2$ and $H_3$. 
As described in Sect. \ref{recog_model}, from the physical knowledge, two recognition models $H_2$ and $H_3$ are expected to be selected for the emulated measurement data of La$_2$O$_3$ and CeO$_2$ XPS spectra, respectively. 
The spectrum of CeO$_2$ has a three-peak structure, and La$_2$O$_3$ has a two-peak structure. 
If the selection of an effective model is based only on the peak number, the effective model could also be selected indirectly by the existing spectral deconvolution method which has no internal model\cite{Nagata12,tokuda16}.  
On the other hand, because our proposed method builds an effective model into spectral deconvolution, various information about peak structure, such as the peak position or the order of peak intensity, suppose to be used to select the effective model. 
To confirm this, we applied our proposed method not only to spectra of CeO$_2$ and La$_2$O$_3$, 
but also to spectra that are the intermediate states of a seamless transition from the CeO$_2$ spectrum to the La$_2$O$_3$ spectrum. 
As we mentioned in Sect. \ref{gen_model}, 
$\Delta$ is an important parameter for controlling the properties of 
XPS spectra from La$_2$O$_3$ to CeO$_2$. 
The generated parameters $V$ and $U_{fc}$ of the La$_2$O$_3$ spectrum are also slightly different from the generated parameters of the CeO$_2$ spectrum (Table \ref{table1}). 
However, the emulated measurement data of the La$_2$O$_3$ spectrum 
generated by replacing the $V$ and $U_{fc}$ values with the parameter values of the CeO$_2$ spectrum 
have almost the same peak position and peak intensity as those in the La$_2$O$_3$ spectrum [Figs. \ref{fig_spectrals}(a)-(1), \ref{fig_spectrals}(a)-(2), and \ref{fig_spectraldata}(a)]. 
Therefore, we refer to this parameter-replaced emulated spectrum as the La$_2$O$_3$ spectrum. 
That is, by shifting $\Delta$ from 1.6 to 12.5 and fixing the other parameters as the parameters of CeO$_2$, 
we can generate spectra that have the intermediate structure of the La$_2$O$_3$ and CeO$_2$ spectra. 
%
The increase in the parameter $\Delta$ from CeO$_2$ to La$_2$O$_3$ induces 
the decrease in the transition probability $\langle{}F_{max}|a_c|G\rangle{}$ from the initial ground state $|G\rangle{}$ 
to the final maximum eigenenergy $E_{max}$ state $<F_{max}|$. 
Because the square of the transition probability $|\langle{}F_{max}|a_c|G\rangle{}|^2$ is the peak intensity, 
the decrease in the transition probability $\langle{}F_{max}|a_c|G\rangle{}$ transforms the three peak CeO$_2$ spectrum 
to a two-peak spectrum. 
To be precise, the emulated measurement data of La$_2$O$_3$ 
has three peaks, but the peak intensity corresponding to the largest eigenvalue $E_2$ 
is very small. 
Hence, in this study, we define the peak number of the emulated measurement data 
as the number of peaks whose intensity is larger than the noise intensity $\sigma_{noise}$. 
In this way, we generated emulated measurement data by shifting the parameter $\Delta$ and applying our proposed method to it. 
More specifically, all parameters except $\Delta$ are the same, 
$\left[V=0.76,U_{ff}=10.5,U_{fc}=12.5,\Gamma = 0.5\right]$, 
for all applied emulated measurement data. 
As mentioned above, the peak number is defined using the noise intensity $\sigma_{noise}$ 
and the peak intensity. 
Therefore, we evaluated the effect of not only $\Delta$, but also 
the noise intensity $\sigma_{noise}$. 
Hence, we also generated the 
emulated measurement data by setting the noise standard deviation to 
$\sigma_{noise} \in \{0.001 ,  0.0028,  0.0046,  0.0064,  0.0082,  0.01 \}$. 
The peak number, peak position, and peak intensity of the emulated measurement data 
are described in Table \ref{table_spectral} and some examples of emulated measurement data 
are described in Fig. \ref{fig_spectrals}(a). 
In the emulated measurement data of $\Delta = 10.08$, if 
the noise intensity $\sigma_{noise}$ is smaller than 0.0052, the peak number is three, 
whereas if the noise intensity $\sigma_{noise}$ is larger than 0.0052, the peak number is two (Table \ref{table_spectral}). 
In this study, each emulated measurement data consists of $N = 400$ samples. \par
To apply Bayesian estimation, we set the prior density ${\rm P}(\boldsymbol{\theta}_k|H_k)$ 
to a uniform distribution in the range described in Table \ref{table2}. 
In the execution of EMC sampling, we adopted the Metropolis\textendash Hastings algorithm\cite{hastings70} to sample each state of inverse temperature. 
The states of inverse temperature 
were determined following the exponential function\cite{Nagata08}:
\begin{eqnarray}
\beta_l = \left\{ \begin{array}{ll}
0.0 & (l=1) \\
\gamma^{l-L} & (l\neq1) \\
\end{array} ,\right.
\end{eqnarray}
where $L = 40$ and $\gamma=1.4$. 
We abandoned the first 10,000 steps and sampled the next 1,000,000 steps. 
\par
On the basis of the obtained Bayesian free energies $F(H_2)$ and $F(H_3)$, 
which correspond to the two-state Hamiltonian model $H_2$ and the three-state Hamiltonian model $H_3$, 
respectively, the log likelihood of $H_3$, ${\rm P}(H_3|\boldsymbol{D})$, was 
calculated as 
\begin{equation}
{\rm P}(H_3|\boldsymbol{D}) = \frac{\exp[-F(H_3)]}{\exp[-F(H_2)]+\exp[-F(H_3)]}.
\end{equation}

If ${\rm P}(H_3|\boldsymbol{D})>0.5$, 
then $H_3$ is a more plausible model than $H_2$. 
Otherwise, if ${\rm P}(H_3|\boldsymbol{D})<0.5$, 
then $H_2$ is a more plausible model than $H_3$. 
From the phase diagram of ${\rm P}(H_3|\boldsymbol{D})$ (Fig. \ref{fig_heatmap}), the three-state Hamiltonian model $H_3$ was selected for the emulated measurement data of 
$\Delta = 1.6$, corresponding to the CeO$_2$ spectra, and the two-state Hamiltonian model $H_2$ was selected for the emulated measurement data 
of $\Delta = 12.5$, corresponding to the La$_2$O$_3$ spectra. 
Furthermore, the proposed method selected the three-state Hamiltonian model $H_3$ for all intermediated  emulated measurement data from $\Delta = 1.6$ to $\Delta = 12.5$. 
It included the emulated measurement data of $\Delta = 10.08$ and $\sigma_{noise}\geq 0.0046$, whose peak intensity of the largest energy was smaller than the noise intensity. 
For further analysis, we evaluated the ratio of the $FE$s, $F(H_3)/F(H_2)$, which corresponds to the difference in likelihoods, 
${\rm P}(H_3|\boldsymbol{D}) - {\rm P}(H_2|\boldsymbol{D})$, in logarithm space. 
$F(H_3)/F(H_2) < 1$ means that $H_3$ is a better model than $H_2$, 
and $F(H_3)/F(H_2) > 1$ means that $H_2$  is a better model than $H_3$. 
The value of $F(H_3)/F(H_2)$ tends to gradually increase as 
the data generated by the parameters $\Delta$ and $\sigma_{noise}$ increase [Fig. \ref{fig_heatmap}(b)]. 
\par
To estimate the model parameters and their estimation uncertainty on the basis of the effective model $H_3$, 
we evaluated the posterior density of parameters ${\rm P}(\boldsymbol{\theta}_3|\boldsymbol{\mathcal{I}})$. 
The posterior density must consist of
independent samples. However, the sampling time series of the EMC 
has a time correlation. 
Therefore, we extracted samples from sufficiently separated 
intervals with a correlation coefficient of 0.5 or lower, 
and calculated the posterior distribution. 
Furthermore, to visualize a posterior distribution 
with more than three dimensions, 
we calculated the following marginal posterior density, 
which marginalizes the posterior distribution 
of the parameters $\boldsymbol{\theta}_k^{\lnot m}$ 
except for the parameter of interest $\theta_k^m$: 
\begin{equation}
{\rm  P}(\theta_k^m|\boldsymbol{\mathcal{I}})  = 
\int_{-\infty}^{\infty} {\rm  P}(\boldsymbol{\theta}_k|\boldsymbol{\mathcal{I}}){\rm  P}(\boldsymbol{\theta}_k)  d\boldsymbol{\theta}_k^{\lnot m}.
\end{equation}
Assuming that the $T$ sampling data of one parameter $m$, $\{\theta_k^{m}(t)\}_{t=1}^T$, 
are extracted from the sampling result of the EMC, 
the marginalized posterior distribution ${\rm  P}(\theta_k^m|\boldsymbol{\mathcal{I}})$ can be estimated from sampling result by kernel density estimation method using Gaussian kernels. 
We determined the bandwidth of Gaussian kernels using Scott's rule\cite{scott 15}. 
\par
From the evaluation of the marginal posterior density of 
the emulated measurement data of $\Delta$ equal to 7.66 or less, 
we found a sharp peak in posterior distributions around the true parameter, which we used to generate the emulated measurement data (Figs. \ref{fig_post_delta} -- \ref{fig_post_gamma3} ). 
From the evaluation of 
the marginal posterior density with the emulated measurement data of $\Delta=10.08$, whose $\Delta$ value is the model estimation switching boundary from $H_3$ to $H_2$, 
we found that the width of the posterior distribution of $\Delta$, $U_{fc}$, and $U_{ff}$ increased 
as the noise intensity $\sigma_{noise}$ (Figs. \ref{fig_post_delta}, \ref{fig_post_ufc}, and \ref{fig_post_uff}) increased.  
On the other hand, the marginal posterior densities of $\Delta, U_{ff}$, and $U_{fc}$ for the emulated measurement data of $\Delta=12.5$  have almost uniform distributions. \par
In this study, model parameters were estimated 
from such posterior distributions 
using the following two methods. 
The first method was the maximum a posteriori (MAP) method. 
The MAP method estimates parameters 
on the basis of the following equation: 
\begin{equation}
\boldsymbol{\theta}_k^{\:MAP} = \argmax_{\boldsymbol{\theta}_k} {\rm  P}(\boldsymbol{\theta}_k|\boldsymbol{\mathcal{I}}).
\end{equation}
The second method was the maximizer of the posterior marginal (MPM) method. 
The MPM method estimates parameters 
on the basis of the following equation: 
\begin{equation}
\theta_k^{m\:MPM} = \argmax_{\theta_k^m} {\rm  P}(\theta_k^m|\boldsymbol{\mathcal{I}}) 
 = \argmax_{\theta_k^m} \int_{-\infty}^{\infty} {\rm  P}(\boldsymbol{\theta}_k|\boldsymbol{\mathcal{I}}){\rm  P}(\boldsymbol{\theta}_k) d\boldsymbol{\theta}_k^{\lnot m},
\end{equation}
where, as with the marginal posterior density, 
$m$ is the index of a certain parameter $\theta_k^m$ 
included in the parameter set $\boldsymbol{\theta}_k$. 
The MPM corresponds to using the maximum marginal posterior density as an estimation value. 
To evaluate the estimation uncertainty of a parameter, 
we defined the variation $\chi^m$ of 
sampling data $\{\theta_k^{m}(t)\}_{t=1}^T$ from the MPM as
\begin{equation}
\chi^m = \frac{1}{T}\sum_{t=1}^T (\theta_k^{m}(t) - \theta_k^{m\:MPM}).
\end{equation}
Here, we focused on the parameters of 
the effective model $H_3$, 
which are $\Delta$, $U_{fc}$, $U_{ff}$, $V$, $\Gamma_2$, and $\Gamma_3$ (Tables \ref{table3}, \ref{table4}, \ref{table5}, \ref{table6}, \ref{table7}, and \ref{table8}, respectively).\par
In the emulated measurement data of $\Delta<$10.08, 
all parameters were estimated correctly 
by both the MAP and MPM methods. 
In greater detail, the gaps between the estimated parameters and the true parameters increased 
as the noise variance increased, and the variation $\chi$ also increased as the noise intensity $\sigma_{noise}$ increased. 
In the emulated measurement data of $\Delta\geq10.08$, 
the gaps between the estimated parameters and the true parameters 
were much larger than the others, except for the emulated 
measurement data of 
$\Delta=10.08$ and $\sigma_{noise}=0.001$ and 0.0028. 
We evaluated the uncertainty of parameter estimation from the variation $\chi$ of the marginal posterior density. 
As a result, at $\Delta=10.08$ and $\sigma_{noise}=0.0046$, 
a large transition of the variation $\chi$ of $\Delta$, $U_{fc}$, and $U_{ff}$ greater than one order was observed (Tables \ref{table3}, \ref{table4}, and \ref{table5}). 
Also, at $\Delta = 12.5$, a large $\chi$ was observed regardless of the noise intensity (Tables \ref{table3}, \ref{table4}, and \ref{table5}). 

\section{Discussion}
By applying our proposed method to emulated measurement data, 
we determined that the two-state Hamiltonian model $H_2$ should be applied to emulated measurement data corresponding to the La$_2$O$_3$ spectra. 
On the other hand, the three-state Hamiltonian model $H_3$ should be applied to emulated measurement data corresponding to the CeO$_2$ spectra. 
These results are consistent with those of previous studies\cite{Kotani74,Kotani85}. 
For the emulated measurement data of $\Delta=10.08$ and $\sigma_{noise} \geq 0.0046$, 
our proposed method selected 
the three-state Hamiltonian model $H_3$. 
These spectral distributions, the same as the La$_2$O$_3$ spectra, are two-peak spectra 
as described in Fig. \ref{fig_spectrals}[a-(2)] and Table \ref{table_spectral}. 
Here, we consider applying the existing method to 
the emulated measurement data of $\Delta=10.08$. 
The existing Bayesian spectral deconvolution methods that are applicable to the analysis of the core-level 3$d$ XPS spectrum have no internal model\cite{Nagata12,tokuda16}. 
Such existing methods simply select the model 
whose number of peaks is the same as the 
appearance of peak numbernumber of peaks appearing in the spectra\cite{tokuda_Dthesis16}. 
Therefore, if the existing methods are applied to the emulated measurement data of $\Delta=10.08$, 
a model that has a two peak structure should be selected. 
Such differences in model selection results between our proposed method and existing Bayesian spectral deconvolution methods will depend on whether the internal model, the effective Hamiltonian, was built in the spectral deconvolution model. 
This suggests that our proposed method should be applied 
to the analysis of core-shell XPS spectra 
when its candidates of the effective Hamiltonian are given. 
\par
From the posterior distribution, 
we can estimate the uncertainties of estimated parameters. 
Actually, from the analysis of the posterior distribution of the $H_3$ model, 
it is confirmed that the estimation uncertainty, 
which corresponds to the variation $\chi$, is increased as the noise intensity 
$\sigma_{noise}$ increased. 
The marginalized posterior distributions, 
except for $\Gamma$ and $V$, of $\Delta=10.08$ 
and $\sigma_{noise}=0.0046$ were broad and had no sharp peak structures, whereas the marginalized posterior distributions of $\Delta=10.08$ and $\sigma_{noise}=0.0028$ had sharp peak structures. 
These properties suggest that the estimation uncertainty significantly decreases around $\Delta=10.08$ and $\sigma_{noise}=0.0028$, 
where the peak number of the spectrum is changed. 
This suggests that, to obtain high estimation accuracy, 
the noise intensity must be reduced to less than $\sigma_{noise}=0.0028$\cite{tokuda_Dthesis16}. 
Through such an analysis of the posterior distribution, 
it is possible to make a measurement plan, 
such as the number of measurements and the measurement method, to satisfy the required estimation accuracy\cite{tokuda_Dthesis16}. 
Such an expansion of the variation $\chi$ of 
the marginalized posterior distribution is presumed 
to occur via the indefinite estimation parameter 
as a result of the effective model $H_3$ having 
an excessive expression power. 
In particular, the fact that the marginalized posterior density began to spread at $\Delta=10.08$ 
and $\sigma_{noise}=0.0046$, where the peak number was changed, suggests that the effective model $H_3$ has excessive expression capability for the two-peak spectrum. 
Information about the estimation accuracy of parameters or the expression capability of the effective model is difficult to obtain by the conventional methods of analysis such as the core-level XPS analysis method using manual tuning and spectral deconvolution using a simple fitting method.\par
In this study, we adopted the simplified cluster model as the effective model of core-level 3$d$ XPS spectra. 
This effective model does not take into account the spin-orbit interaction, the multiplet effect, or the crystal field effect. 
It is generally too simplistic to explain the actual measurement spectra of core-level XPS. 
However, if the spectrum intensity model is generated by Fermi's golden rule, our proposed method can easily replace 
the internal model with another model. 
For example, except for the difficulty related to increasing the number of parameters, 
the effective Hamiltonian used in this study can be easily replaced with a model that takes into account the spin-orbit interaction, the multiplet effect, and the crystal field effect. 
This capability means that we can apply our proposed method to a wider range of actual observed XPS spectra by replacing the effective model with a cluster model that focuses more on interactions or with the impurity Anderson model. 
In the impurity Anderson model, the effect of the band structure of conducting electrons is concerned with the band structure of conducting electrons. 
Thus, it is suggested that the framework of the proposed method has 
wide applicability to actual measurement data of core-level XPS. 
Also, the analysis of emulated measurement data in the intermediate state can also be realized by the analysis of the actual spectra which have the same kind of spectral structure for each other. 
\section{Summary}
By incorporating the effective Hamiltonian into the stochastic model of spectral deconvolution, 
we developed a Bayesian spectral deconvolution method for core-level XPS to realize the automatic analysis of core-level XPS spectra. 
By applying our proposed method to the emulated 3$d$ core-level XPS spectra of La$_2$O$_3$ and CeO$_2$, it was confirmed that our proposed method selects effective Hamiltonians that are consistent with knowledge obtained from the conventional study of physics. 
We also applied our proposed method to spectra which are the intermediate states of a seamless transition from the CeO$_2$ spectrum (three peaks) to the La$_2$O$_3$ spectrum (two peaks). 
As a result, it was confirmed that the proposed method selects an effective Hamiltonian on the basis of not only the information about the peak number but also other information contained in the effective Hamiltonian. 
This cannot be realized by existing Bayesian spectral deconvolution methods applicable to XPS spectra, which select the model on the basis of the peak number\cite{Nagata12, tokuda16}. 
Our proposed method also enables the parameter estimation 
of the effective model using the posterior distribution of its parameter. 
Using the MAP or MPM methods, 
we were able to estimate the true parameters of the generative model 
$\mathcal{H}$ from the posterior distributions. 
Furthermore, using the posterior distribution, 
we were able to evaluate the parameter estimation accuracy 
or obtain information about the properties of the effective model for the spectrum. 
This capability of our proposed method can yield information for scientific discussion, e.g., detection limit\cite{tokuda_Dthesis16}, or the improvement of the effective Hamiltonian using the observed data. 
In conventional analysis methods, such as those using manual tuning or a simple fitting technique, such information cannot be obtained. It is also suggested through our discussion that the framework of the proposed method has wide applicability to actual measurement data of core-level XPS. 
We expect that our proposed method will pave the way for the highly quantitative analysis of core-level XPS spectra. 

{\bf Acknowledgments} This work was supported by the Cross-ministerial Strategic Innovation Promotion Program (SIP), ``Structural Materials for Innovation" (Funding agency: JST) and JST CREST (JPMJCR1761, JPMJCR1861).

\begin{table*}[b]
  \caption{Estimated values of $\Delta$ by MAP and MPM methods. The estimation accuracy can be evaluated by comparison between the estimated parameter and true parameter. And the uncertainty of estimation can be evaluated by the variation $\chi$.}
\begin{center}
  \begin{tabular}{c||c|c|c|c|c|c|c|c|c|c|c|c|c} \hline
$\Delta$\textbackslash$\sigma_{noise}$&&\multicolumn{2}{|c|}{0.001}&\multicolumn{2}{|c|}{0.0028}&\multicolumn{2}{|c|}{0.0046}&\multicolumn{2}{|c|}{0.0064}&\multicolumn{2}{|c|}{0.0082}&\multicolumn{2}{|c}{0.01}\\ \hline  \hline
&True&MAP&MPM$\:\pm\:\chi$&MAP&MPM$\:\pm\:\chi$&MAP&MPM$\:\pm\:\chi$&MAP&MPM$\:\pm\:\chi$&MAP&MPM$\:\pm\:\chi$&MAP&MPM$\:\pm\:\chi$\\ \hline
12.5&12.5&14.68&13.65$ \pm $34.53&18.56&18.97$ \pm $94.66&8.81&10.77$ \pm $30.62&16.07&18.16$ \pm $77.65&18.71&17.89$ \pm $77.3&15.68&16.32$ \pm $58.92\\ \hline
10.08&10.08&8.85&8.93$ \pm $0.28&8.72&8.41$ \pm $8.92&7.94&7.21$ \pm $27.91&4.64&6.65$ \pm $33.98&5.91&9.42$ \pm $24.6&2.33&6.2$ \pm $39.8\\ \hline
7.66&7.66&7.62&7.64$ \pm $0.0&7.58&7.62$ \pm $0.0&7.62&7.58$ \pm $0.01&7.65&7.59$ \pm $0.03&7.32&7.56$ \pm $0.05&7.48&7.58$ \pm $0.07\\ \hline
5.23&5.23&5.22&5.21$ \pm $0.0&5.15&5.17$ \pm $0.0&5.1&5.14$ \pm $0.01&5.13&5.1$ \pm $0.01&5.04&5.05$ \pm $0.02&5.02&5.02$ \pm $0.03\\ \hline
2.81&2.81&2.82&2.81$ \pm $0.0&2.8&2.81$ \pm $0.0&2.79&2.8$ \pm $0.0&2.82&2.8$ \pm $0.0&2.75&2.81$ \pm $0.01&2.71&2.81$ \pm $0.01\\ \hline
1.6&1.6&1.6&1.6$ \pm $0.0&1.62&1.59$ \pm $0.0&1.6&1.6$ \pm $0.0&1.56&1.6$ \pm $0.0&1.61&1.6$ \pm $0.0&1.63&1.59$ \pm $0.0\\ \hline
  \end{tabular}
  \label{table3}
\end{center}
\end{table*}

\begin{table*}[b]
  \caption{Estimated values of $U_{fc}$ by MAP and MPM methods. The estimation accuracy can be evaluated by comparison between the estimated parameter and true parameter. And the uncertainty of estimation can be evaluated by the variation $\chi$.}
\begin{center}
  \begin{tabular}{c||c|c|c|c|c|c|c|c|c|c|c|c|c} \hline
$\Delta$\textbackslash$\sigma_{noise}$&&\multicolumn{2}{|c|}{0.001}&\multicolumn{2}{|c|}{0.0028}&\multicolumn{2}{|c|}{0.0046}&\multicolumn{2}{|c|}{0.0064}&\multicolumn{2}{|c|}{0.0082}&\multicolumn{2}{|c}{0.01}\\ \hline  \hline
&True&MAP&MPM$\:\pm\:\chi$&MAP&MPM$\:\pm\:\chi$&MAP&MPM$\:\pm\:\chi$&MAP&MPM$\:\pm\:\chi$&MAP&MPM$\:\pm\:\chi$&MAP&MPM$\:\pm\:\chi$\\ \hline
12.5&12.5&14.31&5.82$\pm$53.4&18.0&5.78$\pm$55.52&9.68&5.81$\pm$52.92&15.92&5.93$\pm$56.29&17.91&5.92$\pm$54.12&15.33&5.99$\pm$51.92\\ \hline
10.08&12.5&11.61&11.66$\pm$0.14&11.53&11.32$\pm$5.58&10.9&10.68$\pm$17.29&8.79&9.85$\pm$23.41&9.54&9.66$\pm$25.75&7.6&9.73$\pm$25.67\\ \hline
7.66&12.5&12.46&12.48$\pm$0.0&12.43&12.46$\pm$0.0&12.48&12.43$\pm$0.01&12.46&12.44$\pm$0.02&12.19&12.4$\pm$0.03&12.35&12.39$\pm$0.05\\ \hline
5.23&12.5&12.48&12.49$\pm$0.0&12.45&12.47$\pm$0.0&12.42&12.45$\pm$0.0&12.42&12.43$\pm$0.01&12.41&12.4$\pm$0.01&12.39&12.39$\pm$0.01\\ \hline
2.81&12.5&12.5&12.5$\pm$0.0&12.49&12.5$\pm$0.0&12.47&12.49$\pm$0.0&12.49&12.5$\pm$0.0&12.46&12.49$\pm$0.01&12.51&12.48$\pm$0.01\\ \hline
1.6&12.5&12.48&12.49$\pm$0.0&12.51&12.48$\pm$0.0&12.43&12.47$\pm$0.0&12.39&12.48$\pm$0.01&12.37&12.43$\pm$0.01&12.45&12.42$\pm$0.01\\ \hline
  \end{tabular}
  \label{table4}
\end{center}
\end{table*}

\begin{table*}[b]
  \caption{Estimated values of $U_{ff}$ by MAP and MPM methods. The estimation accuracy can be evaluated by comparison between the estimated parameter and true parameter. And the uncertainty of estimation can be evaluated by the variation $\chi$.}
\begin{center}
  \begin{tabular}{c||c|c|c|c|c|c|c|c|c|c|c|c|c} \hline
$\Delta$\textbackslash$\sigma_{noise}$&&\multicolumn{2}{|c|}{0.001}&\multicolumn{2}{|c|}{0.0028}&\multicolumn{2}{|c|}{0.0046}&\multicolumn{2}{|c|}{0.0064}&\multicolumn{2}{|c|}{0.0082}&\multicolumn{2}{|c}{0.01}\\ \hline  \hline
&True&MAP&MPM$\:\pm\:\chi$&MAP&MPM$\:\pm\:\chi$&MAP&MPM$\:\pm\:\chi$&MAP&MPM$\:\pm\:\chi$&MAP&MPM$\:\pm\:\chi$&MAP&MPM$\:\pm\:\chi$\\ \hline
12.5&10.5&9.75&13.43$\pm$11.0&12.41&14.91$\pm$10.46&11.83&18.66$\pm$31.3&15.57&14.52$\pm$13.12&8.93&14.71$\pm$15.13&11.71&13.37$\pm$14.95\\ \hline
10.08&10.5&11.13&11.05$\pm$0.13&11.32&11.41$\pm$7.87&10.68&15.17$\pm$8.03&12.34&17.38$\pm$11.87&11.1&18.73$\pm$20.31&13.67&18.93$\pm$22.6\\ \hline
7.66&10.5&10.52&10.52$\pm$0.0&10.6&10.54$\pm$0.01&10.62&10.55$\pm$0.01&10.5&10.56$\pm$0.03&10.64&10.58$\pm$0.05&10.58&10.61$\pm$0.08\\ \hline
5.23&10.5&10.51&10.52$\pm$0.0&10.57&10.56$\pm$0.0&10.59&10.58$\pm$0.01&10.56&10.62$\pm$0.01&10.69&10.66$\pm$0.02&10.71&10.68$\pm$0.03\\ \hline
2.81&10.5&10.5&10.5$\pm$0.0&10.51&10.49$\pm$0.01&10.51&10.53$\pm$0.02&10.47&10.49$\pm$0.04&10.58&10.54$\pm$0.06&10.83&10.51$\pm$0.09\\ \hline
1.6&10.5&10.46&10.48$\pm$0.0&10.47&10.46$\pm$0.01&10.3&10.39$\pm$0.02&10.28&10.38$\pm$0.03&10.09&10.39$\pm$0.05&10.25&10.28$\pm$0.08\\ \hline
  \end{tabular}
  \label{table5}
\end{center}
\end{table*}

\begin{table*}[b]
  \caption{Estimated values of $V$ by MAP and MPM methods. The estimation accuracy can be evaluated by comparison between the estimated parameter and true parameter. And the uncertainty of estimation can be evaluated by the variation $\chi$.}
\begin{center}
  \begin{tabular}{c||c|c|c|c|c|c|c|c|c|c|c|c|c} \hline
$\Delta$\textbackslash$\sigma_{noise}$&&\multicolumn{2}{|c|}{0.001}&\multicolumn{2}{|c|}{0.0028}&\multicolumn{2}{|c|}{0.0046}&\multicolumn{2}{|c|}{0.0064}&\multicolumn{2}{|c|}{0.0082}&\multicolumn{2}{|c}{0.01}\\ \hline  \hline
&True&MAP&MPM$\:\pm\:\chi$&MAP&MPM$\:\pm\:\chi$&MAP&MPM$\:\pm\:\chi$&MAP&MPM$\:\pm\:\chi$&MAP&MPM$\:\pm\:\chi$&MAP&MPM$\:\pm\:\chi$\\ \hline
12.5&0.76&0.77&0.75$\pm$0.02&0.76&0.75$\pm$0.02&0.72&0.75$\pm$0.02&0.75&0.75$\pm$0.01&0.78&0.75$\pm$0.02&0.76&0.75$\pm$0.02\\ \hline
10.08&0.76&0.73&0.73$\pm$0.00&0.72&0.72$\pm$0.01&0.71&0.8$\pm$0.03&0.58&0.8$\pm$0.03&0.64&0.8$\pm$0.03&0.42&0.8$\pm$0.03\\ \hline
7.66&0.76&0.76&0.76$\pm$0.00&0.76&0.76$\pm$0.00&0.76&0.76$\pm$0.00&0.76&0.76$\pm$0.00&0.76&0.76$\pm$0.00&0.75&0.76$\pm$0.00\\ \hline
5.23&0.76&0.76&0.76$\pm$0.00&0.76&0.76$\pm$0.00&0.76&0.76$\pm$0.00&0.76&0.76$\pm$0.00&0.75&0.75$\pm$0.00&0.75&0.75$\pm$0.00\\ \hline
2.81&0.76&0.76&0.76$\pm$0.00&0.76&0.76$\pm$0.00&0.76&0.76$\pm$0.00&0.76&0.76$\pm$0.00&0.76&0.76$\pm$0.00&0.76&0.76$\pm$0.00\\ \hline
1.6&0.76&0.76&0.76$\pm$0.00&0.76&0.76$\pm$0.00&0.76&0.76$\pm$0.00&0.76&0.76$\pm$0.00&0.76&0.76$\pm$0.00&0.76&0.76$\pm$0.00\\ \hline
  \end{tabular}
  \label{table6}
\end{center}
\end{table*}

\begin{table*}[b]
  \caption{Estimated values of $\Gamma_2$ by MAP and MPM methods. The estimation accuracy can be evaluated by comparison between the estimated parameter and true parameter. And the uncertainty of estimation can be evaluated by the variation $\chi$.}
\begin{center}
  \begin{tabular}{c||c|c|c|c|c|c|c|c|c|c|c|c|c} \hline
$\Delta$\textbackslash$\sigma_{noise}$&&\multicolumn{2}{|c|}{0.001}&\multicolumn{2}{|c|}{0.0028}&\multicolumn{2}{|c|}{0.0046}&\multicolumn{2}{|c|}{0.0064}&\multicolumn{2}{|c|}{0.0082}&\multicolumn{2}{|c}{0.01}\\ \hline  \hline
&True&MAP&MPM$\:\pm\:\chi$&MAP&MPM$\:\pm\:\chi$&MAP&MPM$\:\pm\:\chi$&MAP&MPM$\:\pm\:\chi$&MAP&MPM$\:\pm\:\chi$&MAP&MPM$\:\pm\:\chi$\\ \hline
12.5&0.5&0.5&0.5$\pm$0.00&0.5&0.5$\pm$0.00&0.51&0.5$\pm$0.00&0.5&0.5$\pm$0.00&0.5&0.51$\pm$0.00&0.5&0.51$\pm$0.00\\ \hline
10.08&0.5&0.5&0.5$\pm$0.00&0.5&0.5$\pm$0.00&0.49&0.5$\pm$0.00&0.48&0.49$\pm$0.00&0.49&0.49$\pm$0.00&0.5&0.49$\pm$0.00\\ \hline
7.66&0.5&0.5&0.5$\pm$0.00&0.5&0.5$\pm$0.00&0.5&0.5$\pm$0.00&0.5&0.5$\pm$0.00&0.5&0.5$\pm$0.00&0.5&0.5$\pm$0.00\\ \hline
5.23&0.5&0.5&0.5$\pm$0.00&0.5&0.5$\pm$0.00&0.5&0.5$\pm$0.00&0.51&0.5$\pm$0.00&0.51&0.5$\pm$0.00&0.5&0.5$\pm$0.00\\ \hline
2.81&0.5&0.5&0.5$\pm$0.00&0.5&0.5$\pm$0.00&0.52&0.51$\pm$0.00&0.51&0.51$\pm$0.00&0.52&0.51$\pm$0.00&0.5&0.52$\pm$0.00\\ \hline
1.6&0.5&0.5&0.5$\pm$0.00&0.5&0.5$\pm$0.00&0.5&0.5$\pm$0.00&0.51&0.5$\pm$0.00&0.51&0.5$\pm$0.00&0.5&0.5$\pm$0.00\\ \hline
  \end{tabular}
  \label{table7}
\end{center}
\end{table*}

\begin{table*}[b]
  \caption{Estimated values of $\Gamma_3$ by MAP and MPM methods. The estimation accuracy can be evaluated by comparison between the estimated parameter and true parameter. And the uncertainty of estimation can be evaluated by the variation $\chi$.}
\begin{center}
  \begin{tabular}{c||c|c|c|c|c|c|c|c|c|c|c|c|c} \hline
$\Delta$\textbackslash$\sigma_{noise}$&&\multicolumn{2}{|c|}{0.001}&\multicolumn{2}{|c|}{0.0028}&\multicolumn{2}{|c|}{0.0046}&\multicolumn{2}{|c|}{0.0064}&\multicolumn{2}{|c|}{0.0082}&\multicolumn{2}{|c}{0.01}\\ \hline  \hline
&True&MAP&MPM$\:\pm\:\chi$&MAP&MPM$\:\pm\:\chi$&MAP&MPM$\:\pm\:\chi$&MAP&MPM$\:\pm\:\chi$&MAP&MPM$\:\pm\:\chi$&MAP&MPM$\:\pm\:\chi$\\ \hline
12.5&0.5&0.23&0.13$\pm$0.21&0.06&0.34$\pm$0.11&0.34&0.63$\pm$0.1&0.95&0.2$\pm$0.18&0.15&0.81$\pm$0.17&0.84&0.69$\pm$0.11\\ \hline
10.08&0.5&0.3&0.32$\pm$0.01&0.21&0.25$\pm$0.1&0.66&0.71$\pm$0.12&0.05&0.93$\pm$0.25&0.05&0.93$\pm$0.25&0.99&0.48$\pm$0.08\\ \hline
7.66&0.5&0.5&0.5$\pm$0.00&0.48&0.5$\pm$0.00&0.49&0.49$\pm$0.00&0.51&0.49$\pm$0.00&0.46&0.49$\pm$0.00&0.47&0.49$\pm$0.01\\ \hline
5.23&0.5&0.5&0.5$\pm$0.00&0.5&0.5$\pm$0.00&0.51&0.5$\pm$0.00&0.5&0.5$\pm$0.00&0.5&0.51$\pm$0.00&0.51&0.51$\pm$0.00\\ \hline
2.81&0.5&0.5&0.5$\pm$0.00&0.5&0.5$\pm$0.00&0.5&0.5$\pm$0.00&0.5&0.5$\pm$0.00&0.48&0.5$\pm$0.00&0.5&0.5$\pm$0.00\\ \hline
1.6&0.5&0.5&0.5$\pm$0.00&0.5&0.5$\pm$0.00&0.49&0.5$\pm$0.00&0.49&0.5$\pm$0.00&0.5&0.49$\pm$0.00&0.51&0.49$\pm$0.00\\ \hline
  \end{tabular}
  \label{table8}
\end{center}
\end{table*}

\end{document}